\documentclass[aps,prx,superscriptaddress,twocolumn]{revtex4-2}
 \usepackage{graphicx,color}
 \usepackage{verbatim}
 \usepackage{amssymb}   
 \usepackage{amsmath}
 \usepackage{amsfonts}
 \usepackage{mathdots}
 \usepackage{hyperref}
 \usepackage{epsfig}
 \usepackage{hyperref}
 \usepackage{xcolor}
 \definecolor{myblue}{RGB}{46, 48,146}
 \hypersetup{colorlinks=true,linkcolor=myblue,citecolor=myblue,urlcolor=myblue, linktocpage}
 \usepackage{braket}
 \usepackage{bm}
 \usepackage{multirow}
 
\begin{document}
 	\title{Sliding Dynamics of Current-Driven Skyrmion Crystal and Helix in Chiral Magnets}
 	
 	\author{Ying-Ming Xie}
    \affiliation{RIKEN Center for Emergent Matter Science (CEMS), Wako, Saitama 351-0198, Japan} 	
    \author{Yizhou Liu}
     \affiliation{RIKEN Center for Emergent Matter Science (CEMS), Wako, Saitama 351-0198, Japan} 
    \author{Naoto Nagaosa}\thanks{nagaosa@riken.jp}
   \affiliation{RIKEN Center for Emergent Matter Science (CEMS), Wako, Saitama 351-0198, Japan} 

\begin{abstract}

The skyrmion crystal (SkX) and helix (HL) phases, present in typical chiral magnets, can each be considered as forms of density waves but with distinct topologies. The SkX exhibits gyrodynamics analogous to electrons under a magnetic field, while the HL state resembles topological trivial spin density waves. However, unlike the charge density waves, the theoretical analysis of the sliding motion of SkX and HL remains unclear, especially regarding the similarities and differences in sliding dynamics between these two spin density waves. In this work, we systematically explore the sliding dynamics of SkX and HL in chiral magnets in the limit of large current density. We demonstrate that the sliding dynamics of both SkX and HL can be unified within the same theoretical framework as density waves, despite their distinct microscopic orders. Furthermore, we highlight the significant role of gyrotropic sliding induced by impurity effects in the SkX state, underscoring the impact of nontrivial topology on the sliding motion of density waves. Our theoretical analysis shows that the effect of impurity pinning is much stronger in HL compared with SkX, i.e., $\chi^{SkX}/\chi^{HL}\sim \alpha^2$ ($\chi^{SkX}$, $\chi^{HL}$: susceptibility to the impurity potential, $\alpha$ ($\ll 1$) is the Gilbert damping). Moreover, the velocity correction is mostly in the transverse direction to the current in SkX.
These results are further substantiated by realistic Landau-Lifshitz-Gilbert simulations.


 	\end{abstract}
 	
 	\date{\today}
 	
 	\maketitle

\emph{Introduction.---} 
Density waves in solids represent a prevalent phenomenon, particularly in low-dimensional systems \cite{RevModPhys.60.1129,RevModPhys.66.1}. They break the translational symmetry of the crystal, leading to the emergence of Goldstone bosons, i.e., phasons, which remain gapless when the period of density waves is incommensurate with the crystal periodicity. The sliding motion of density waves under an electric field $\bm{E}$ has been extensively studied. In this context, the impurity pinning of phasons results in a finite threshold field \cite{RevModPhys.60.1129,RevModPhys.66.1}. In general, exploring the dynamics of pinning and depinning offers valuable insights into understanding the behavior of density waves.

The skyrmion crystal (SkX) and helix (HL) phases in chiral magnets can be recognized as periodic density waves of spins, as depicted in Figs.~\ref{fig:fig1} (a) and (b). The HL phase is stabilized in chiral magnet at small magnetic field regions, with spins of neighboring magnetic moments arranging themselves in a helical pattern. SkX is a superposition of three phase-locked HL and comprises arrays of magnetic skyrmions, nanoscale vortex-like spin textures characterized by a non-zero skyrmion number $N_{sk}=\frac{1}{4\pi}\int\int d^2\bm{r} \bm{s}\cdot(\partial_x \bm{s} \times \partial_y \bm{s} )$ ($\bm{s}$ being the unit vector of spin). Theoretically proposed magnetic skyrmions \cite{bogdanov1989,BOGDANOV1994,Roler2006} were initially observed in the chiral magnet MnSi under magnetic fields \cite{Boni2009, Yu2010,Heinze2011}, wherein the skyrmion lattice structure produces a six-fold neutron scattering pattern.  Since then, the chiral magnetic states encompassing SkX and HL states have been the focus of extensive research \cite{Nagaosa2013,Fert2017,Bogdanov2020,Back_2020,Tokura2021}.

The dynamics of SkX in a random environment, specifically the pinning effects from impurities, are manifested through the topological Hall effect. The current dependence of topological Hall resistivity $\rho_{xy}$ was initially explored theoretically by Zang et al \cite{Jiadong2011} and experimentally by Schulz et al. \cite{Schulz2012}. To illustrate, a schematic plot is presented in Fig.~\ref{fig:fig1}(c). Typically, there are three distinct regions characterizing the dynamics of SkX: the pinned, creep, and flow regions. The topological Hall resistivity decreases when SkX is depinned because the motion of SkX induces temporal changes in the emerging magnetic fields $\bm{B}^{e}$, subsequently generating emergent electric fields $\bm{E}^{e}$ and an opposing Hall contribution. Theoretically, the pinning problem of both SkX and HL was investigated in terms of replica symmetry breaking \cite{PhysRevB.97.024413}, revealing a distinct difference in glassy states between SkX and HL. The key factor lies in the nontrivial topology of SkX, contrasting with the trivial topology in HL and most density wave states. However, this difference has not been theoretically explored in the context of sliding/moving density wave states for chiral magnets.


In this work, we systemically study the current-driven sliding dynamics of the SkX and HL in chiral magnets.   We employ the methodology proposed by Sneddon et al. \cite{Sneddon1982} in their investigation of charge density waves and apply it to magnetic materials. This method allows us to investigate the current-driven dynamics of SkX and HL, considering both deformation and impurity pinning effects. 
  Through this method, we reveal that the drift velocity correction 
$\Delta \bm{v}_d$  due to the impurity pinning effects versus the current density  $j_s$
in the flow region, follows $\Delta \bm{v}_d \propto (v_{d0})^{\frac{d-2}{2}} (-\bm{e}_{\parallel}+  \frac{G}{\alpha D} \bm{e}_{\perp}) $
for the SkX phase, while 
$	\Delta \bm{v}_d\propto -(v_{d0})^{\frac{d-2}{2}} \bm{e}_{\parallel}
	 $ for the HL phase with the spatial dimension denoted as $d$.  Here,   $\bm{e}_{\parallel}$ represents the direction of the intrinsic drift velocity $\bm{v}_{d0}$ (the magnitude of $v_{d0}$ is proportional to the current density $j_s$ due to the universal linear current-velocity relation \cite{Iwasaki2013}), $G= 4 \pi N_{sk}$, $D$ is a form factor   
at the order of unity, $\alpha \ll 1$ is the Gilbert damping parameter so that $G/\alpha D \gg 1$. Although the scaling relation $(v_{d0})^{\frac{d-2}{2}}$ applies to both SkX and HL, we can see that the gyrodynamics of the SkX state induced by its nontrivial topology results in its sliding dynamics more robust than in HL and mostly in the transverse direction. 
Finally, we explicitly conduct the micromagnetic simulations on both the SkX and HL systems, aligning well with our theoretical expectations. 

Our work demonstrates the unification of sliding dynamics between spin density waves and charge density waves within the same theoretical framework. Our results also vividly illuminate both the similarities and differences in the sliding dynamics between SkX and HL phases. This insight significantly enhances our understanding of the sliding dynamics associated with topological density wave phenomena, which possesses possible applications in areas, such as skyrmion-based devices \cite{Jonietz2010, Iwasaknanoi2013,Fert2013}, depinning dynamics \cite{Lin2013,Reichhardt2015,Reichhardt_2016,Kim2017,Legrand2017,Koshibae2018,Reichhardt2020, masell_combing_2020, Reichhardt_RMP}, Hall responses \cite{Jiadong2011,Schulz2012, Tokura2019}, and current-driven motion of Wigner crystals under out-of-plane magnetic fields \cite{Millis1994,Reichhardt2022,Madathil2023}.




 \begin{figure}[t]
 	\centering
 	\includegraphics[width=1\linewidth]{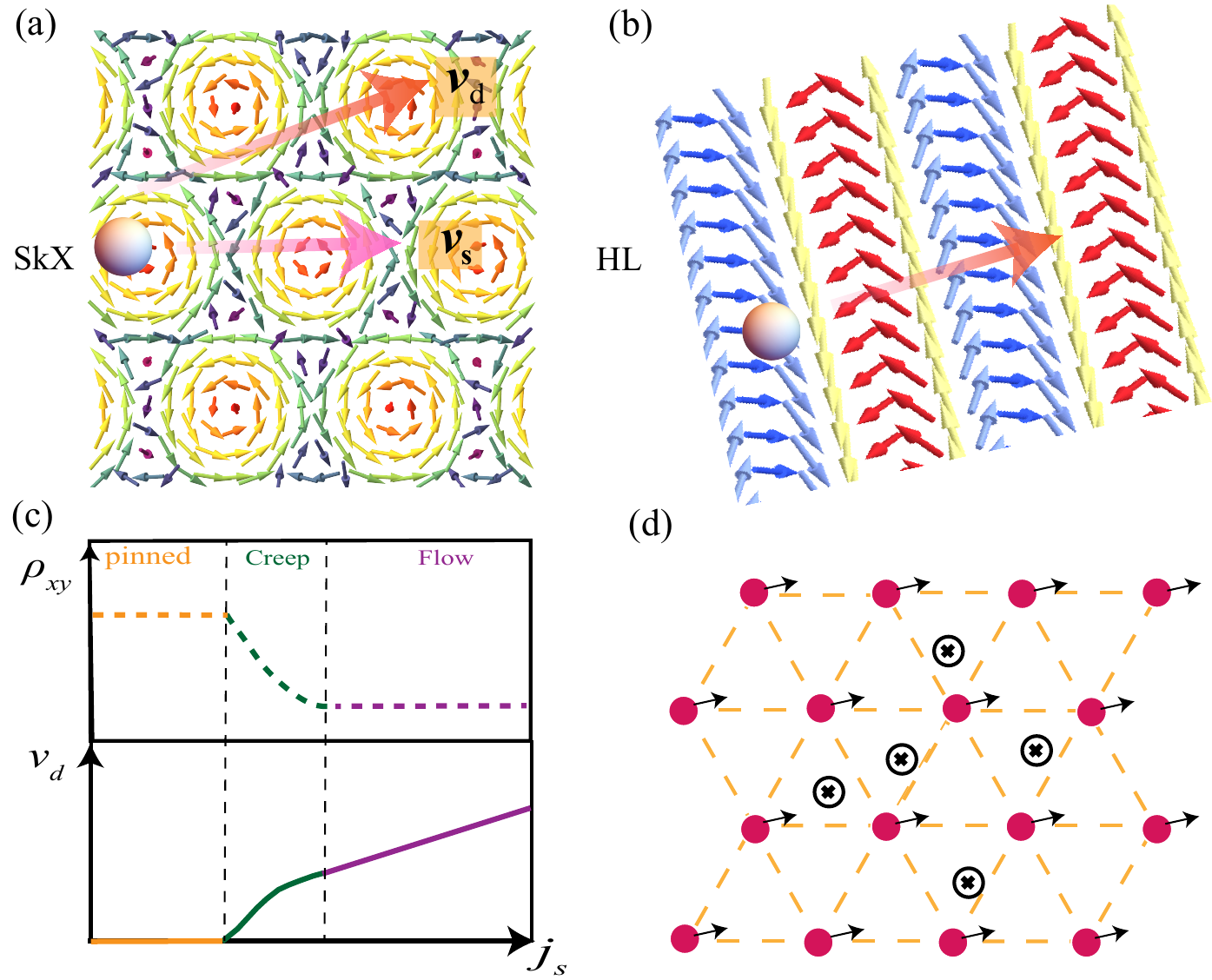}
 	\caption{(a), (b) The current-driven motion of the SkX and HL, respectively. (c) Schematic of the Hall resistivity $\rho_{xy}$ and drift velocity $v_d$ versus current density $\bm{j}_s$ with pinned (yellow), creeping (green), and flowing (purple)  highlighted.  (d) The collective flow motion of the SkX, where the center of each skyrmion (red dots) and the impurities (black crosses) are highlighted.  }
 	\label{fig:fig1}
 \end{figure}

\emph{Sliding dynamics for skyrmion crystals.---}  
The current-driven motion of SkX is described by the Thiele equation assuming that its shape does not change  \cite{Iwasaki2013, Thiele1973, Rosch2012}:
\begin{equation}
\bm{G}\times(\bm{v}_s-\bm{v}_d)+\bm{D}(\beta\bm{v}_s-\alpha\bm{v}_d)+\bm{F}=0.
\end{equation}
Here, the first term on the left represents the Magnus force, the second term is the dissipative force, and the last term arises from the deformation and impurity-pinning effects. Here, $\bm{v}_s$ is the velocity of conduction electrons, $\alpha$ is the damping constant of the magnetic system, and $\beta$ describes the non-adiabatic effects of the spin-polarized current. The gyromagnetic coupling vector is denoted as $\bm{G}=(0,0, 4\pi N_{sk})$, and the dissipation matrix $D_{ij}=\delta_{ij} D$ where $i,j\in \{x,y\}$. It is noteworthy that the Thiele equation respects out-of-plane rotational symmetry [Supplementary Material (SM) Sec. IA \cite{Supp}].

  To obtain the equation of motion of SkX,  the displacement vector field of skyrmions is defined as $\bm{u}(\bm{r},t)$ so that the drift velocity $\bm{v}_d=\frac{\partial \bm{u}(\bm{r},t)}{\partial t}$, where $\bm{r}$ is the position vector, $t$ is the time. The force $\bm{F}$ can be expressed with  $\bm{u}(\bm{r},t)$ as  $	\bm{F}(\bm{r},t)=\bm{F}_{imp}+\bm{F}_{de}$, where the impurity pinning force $\bm{F}_{imp}=	-\sum_{i}\nabla U(\bm{r}+\bm{u}(\bm{r},t)-\bm{r}_i)\rho(\bm{r})=\bm{f}_{imp}(\bm{r}+\bm{u}(\bm{r},t))\rho(\bm{r})$ and the deformation force $	\bm{F}_{de}=	\int d\bm{r'} \mathcal{D}(\bm{r}-\bm{r'})\bm{u}(\bm{r'},t')$. Here, $U(\bm{r}-\bm{r}_i)$ is the impurity potential around site $\bm{r}_i$, $\mathcal{D}(\bm{r}-\bm{r'})$ characterizes the  restoration strength after deformation, and $\rho(\bm{r})$ is the skyrmion density. Based on these definitions, the Thiele equation can be expressed as an equation of motion:
\begin{eqnarray}
	\frac{\partial \bm{u}(\bm{r},t)}{\partial t}&&=	\hat{M}_0\bm{v}_s+
	\hat{M}_1	\int d\bm{r'} \mathcal{D}(\bm{r}-\bm{r'})\bm{u}(\bm{r'},t)\nonumber\\	
	&&+\hat{M}_1\bm{f}_{imp}(\bm{r}+\bm{u}(\bm{r},t))\rho(\bm{r}) .\label{mEq:motion}
\end{eqnarray}
where  $\hat{M}_0=	\frac{1}{G^2+\alpha^2D^2}\begin{pmatrix}
	G^2+\alpha\beta D^2&GD(\beta-\alpha)\\GD(\alpha-\beta)&G^2+\alpha\beta D^2
\end{pmatrix}$ and $\hat{M}_1=\frac{1}{G^2+\alpha^2D^2}\begin{pmatrix}
\alpha D&G\\
-G&\alpha D
\end{pmatrix}$.  Note that each skyrmion is now considered as a center-of-mass particle, and these skyrmions form a triangular lattice and move collectively with scatterings from impurities, as illustrated in Fig.~\ref{fig:fig1}(d).

The displacement vector can be expanded  around the uniform motion,
\begin{equation}
\bm{u}(\bm{r},t)=\bm{v_d}t+\tilde{\bm{u}}(\bm{r},t).
\end{equation} 
Here, $\bm{v_d}$ is the dominant uniform skyrmion motion velocity,  $\tilde{\bm{u}}(\bm{r},t)$ characterizes a small non-uniform part. Using the Green's function approach to solve the differential equation Eq.~\eqref{mEq:motion}, $\tilde{\bm{u}}(\bm{r},t)$ can be obtained as \cite{Sneddon1982,Millis1994,Supp} 
\begin{eqnarray}
		\bm{\tilde{u}}(\bm{r},t)&&=\int d\bm{r'}\int dt' \mathcal{G}(\bm{r}-\bm{r'},t-t')\{\bm{v}_{d0}-\bm{v_d}	\nonumber\\
		&&+M_1\bm{f}_{imp}(\bm{r'}+\bm{v_d}t'+\bm{\tilde{u}}(\bm{r'},t'))\rho(\bm{r'})\},\label{Eq:motion_m}
\end{eqnarray}
where the intrinsic drift velocity $\bm{v}_{d0}=\hat{M}_0\bm{v}_s$,  the Fourier component of the Green's function $\mathcal{G}$ is given by
\begin{equation}
	\mathcal{G}^{-1}(\bm{k},\omega)=-i\omega-\hat{M}_1 \mathcal{D}(\bm{k}). \label{Eq:Green}
\end{equation}
Here, $\mathcal{D}(\bm{k})$ arises from the Fourier transformation of deformation  $\mathcal{D}(\bm{k})=\int d^{d}(\bm{r}) \mathcal{D}(\bm{r}) e^{-i\bm{k}\cdot\bm{r}}$ (the spatial dimension is denoted as $d$).

 In the flow region, $ \tilde{\bm{u}}(\bm{r},t)$ in Eq.~\eqref{Eq:motion_m} can be solved perturbatively. Up to the second order, $\tilde{\bm{u}}(\bm{r},t)\approx \tilde{\bm{u}}_0(\bm{r},t)+\tilde{\bm{u}}_1(\bm{r},t)+\tilde{\bm{u}}_2(\bm{r},t)$, which, respectively, are obtained by replacing the terms in the brackets of Eq.~\eqref{Eq:motion_m}  as $M_0\bm{v_s}-\bm{v}_{\bm{d}}, M_1\bm{f}_{imp}(\bm{r'}+\bm{v}_{d}t')\rho(\bm{r'}), M_1\nabla\bm{f}_{imp}(\bm{r'}+\bm{v}_dt')\cdot \tilde{\bm{u}}_1(\bm{r},t) \rho(\bm{r'})$.  Based on this approximation and  making use of  $\braket{\frac{\tilde{u}(\bm{r},t)}{\partial t}}=0$, the self-consistent equation for the velocity reads (for details see SM Sec.~IB \cite{Supp})
 \begin{eqnarray}
 	&&\bm{v}_{d}=\bm{v}_{d0}+\sum_{\bm{g}}\int \frac{d^d\bm{q}}{(2\pi)^d} |\rho(\bm{g})|^2  \Lambda (\bm{q})\hat{M}_1 \times\nonumber\\
 	&& \begin{pmatrix}
 		q_x^2&q_xq_y\\ q_xq_y&q_y^2
 	\end{pmatrix}\text{Im}\left[\mathcal{G}(\bm{q}-\bm{g},-\bm{q}\cdot\bm{v_d})\right]\hat{M}_1\begin{pmatrix}
 	q_x\\q_y
 	\end{pmatrix}. \label{Eq:velocity}
 \end{eqnarray}
 where   $\rho(\bm{g})$ is the Fourier component of $\rho(\bm{r})$ with $\bm{g}$ as the reciprocal skyrmion lattice vectors, and $\Lambda(\bm{q})$ arises from the impurity average $\overline{U(\bm{q}_1)U(\bm{q}_2)}=(2\pi)^d\Lambda(\bm{q}_2)\delta(\bm{q}_1+\bm{q}_2)$. The impurity strength and functional profile are encoded in $\Lambda(\bm{q})$. Note that the crucial aspects for the above method to be valid are (i) the impurity strength is weak.  (ii) the drift velocity is large compared to the impurity effects and the SkX remains elastic. (iii) the deformation within each skyrmion is negligible so that each skyrmion can be regarded as a point object. Our consideration here is the crystal flowing limit, while those chiral magnetic systems with glass flowing \cite{Reichhardt2015, Reichhardt2020} are not within this scope of the work.

To proceed further, we adopt the following approximations.  The current-driven distortion is expected to be weak so that $\mathcal{D}(\bm{k})$ would be dominant by the long-wave limit. In this case, $\mathcal{D}(\bm{k})$ can be expanded as $K_xk_x^2+K_yk_y^2$ for the 2D case and as $K_xk_x^2+K_yk_y^2+K_zk_z^2$ for the 3D case.   
On the other hand, the characterized frequency that enters into the Green's function is $\bm{q}\cdot \bm{v}_d\sim v_d/a$. Using a reasonable parameter $v_d=10$ m/s, the skyrmion lattice constant  $a=25$ nm, we estimate  $v_d/a \sim 0.4$ GHz. This frequency is much smaller compared with the one of $K_j$, which is roughly the scale of exchange energy $J\sim 1$ meV $\sim 240$ GHz \cite{Iwasaki2013, Jiadong2011}. As a result, the dominant contribution to the integral is given by the elastic modes $v_d g_j\approx \omega_{\bm{k}}\approx  \mathcal{D}(\bm{k})/\sqrt{G^2+\alpha^2D^2}$ with $\bm{k}=\bm{q}-\bm{g} \rightarrow 0$, around which the imaginary part of Green's function is the largest.

With the above approximations, we perform the integral in Eq.~\eqref{Eq:velocity} and sum over the smallest $\bm{g}$ vectors: $\bm{g}_j=\sqrt{3}\kappa_0(\sin \frac{(j-1)\pi}{3}, \cos \frac{(j-1)\pi}{3})$ with $j$ as integers from 1 to 6 and $\kappa_0=\frac{4\pi}{3a}$. Since the Thiele equation exhibits out-of-plane rotational symmetry,  without loss of generality, we set $v_{d0}$ along $x$-direction here. After some simplifications (for details see SM. Sec IB),  we find the correction ($\Delta \bm{v}_d=\bm{v}_d-\bm{v}_{d0}$) on the drift velocity  due to the impurity and deformation are given by
\begin{equation}
	\Delta \bm{v}_d\approx  \chi_{d}^{SkX} (v_{d0})^{\frac{d-2}{2}} (-\bm{e}_{\parallel}+\frac{G}{\alpha D} \bm{e}_{\perp}), \label{Eq:final}
\end{equation} 
where the susceptibility to the impurity potential $\chi_{d}^{SkX}=\frac{9\kappa_0^3|\rho_1|^2\Lambda_0 \alpha D}{4\sqrt{K_xK_y}(G^2+\alpha^2D^2)}$ for $d=2$, while $\chi_{d}^{SkX}=\frac{9\sqrt{3}\kappa_0^{7/2}\Gamma(\frac{G^2+\alpha^2D^2}{4\alpha^2D^2})|\rho_1|^2\Lambda_0(\alpha D)^{3/2}}{\pi^2\sqrt{K_xK_yK_z}(G^2+\alpha^2D^2)}$ for $d=3$ (the function $\Gamma(a)= \int_0^{+\infty} dx \frac{x^6}{(x^4-a)^2+x^4}$). Note that we have replaced $\rho(g_j)=\rho_1,  \Lambda(g_j)=\Lambda_0$ given the six-fold rotational symmetry of the skyrmion lattice. 

The first important aspect in Eq.~\eqref{Eq:final} is that the correction $\Delta v_{d}$ is insensitive to $v_{d0}$ in 2D limit  but follows a square-root scaling: $ (v_{d0})^{1/2}$ in 3D limit. Similar to many scaling phenomena, the dimension plays a critical role here.  The second important aspect is that the correction along the transverse direction directly reflects the skyrmion topological number $G$ with the ratio compared to the longitudinal one as $G/\alpha D$. These interesting aspects embedded in the Eq.~\eqref{Eq:final} will be further highlighted later.

\begin{figure}
	\centering
	\includegraphics[width=1\linewidth]{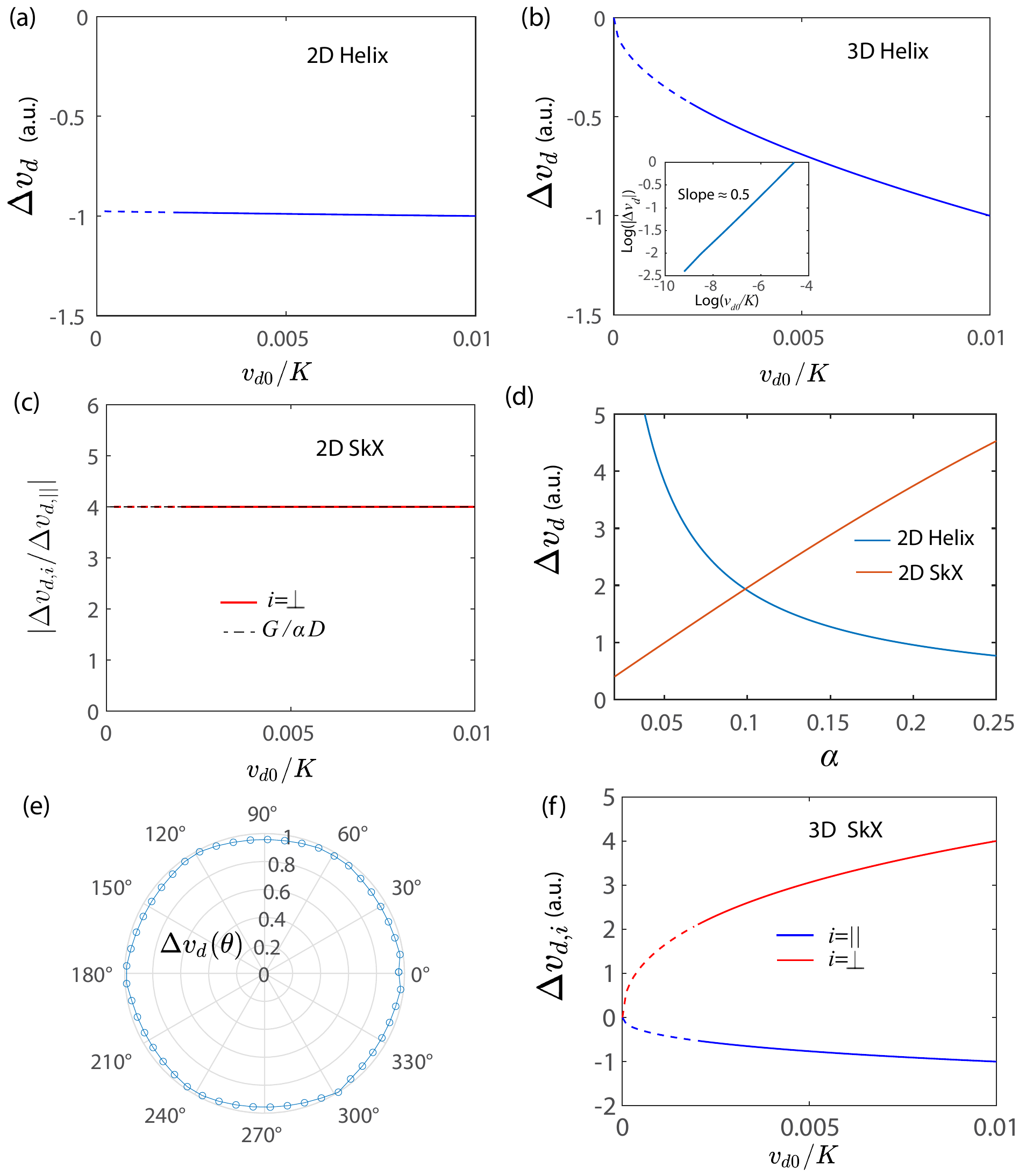}
	\caption{(a) and (b) The correction $\Delta v_{d}$ versus $v_{d0}$ (in units of $K$) for the 2D and 3D HL state, respectively. (c)  shows that the velocity correction ratio between the longitudinal and transverse direction in 2D SkX approaches $\frac{G}{\alpha D}$. The blue and red dashed line regions in (a) and (c) are to highlight the small $v_{d0}$ regions that should be ignored ($v_{d0}=0$ point is not reached). 
 (d) The damping $\alpha$ dependence of the drift velocity correction from the developed model, where the 2D SkX scales as  $\frac{\alpha D}{G^2+\alpha^2D^2}$ and 2D Helix scales as $\frac{1}{\alpha D}$. 
 (e) The angular-dependence of $|\Delta \bm{v}_d(\theta)|$, where $\theta$ is the angle of $\bm{v}_{d0}$. (f) The velocity correction along the longitudinal and transverse directions in the 3D SkX. All the longitudinal components of $\Delta v_{d}$ in these plots have been normalized so the velocity correction is in arbitrary scale (a.u.).  The used parameters are $G=4\pi$, $D=5\pi$, $\alpha=0.2$.  }
	\label{fig:fig2}
\end{figure}
\emph{Helix case.---} It is straightforward to generalize the above treatment to the helical spin order.  The Thiele equation is reduced to one dimension:
 \begin{equation}
 	D(\beta v_s-\alpha v_{d})+F=0. \label{Eq:mo}
 \end{equation}
The essential difference here is the absence of gyrotropic coupling ($G=0$).  Following the same procedure [SM Sec. II], the self-consistent equation for the drift velocity is given by
\begin{equation}
	v_d=	v_{d0}+\int\frac{d^d\bm{q}}{(2\pi)^d}\sum_{\bm{g}}\frac{|\rho(\bm{g})|^2}{\alpha^2D^2 }\Lambda(\bm{q})q_x^3\text{Im}[\mathcal{G}(\bm{q}-\bm{g},-q_xv_d)]. \label{Eq:helical}	
\end{equation} 
Here, $v_{d0}=\frac{\beta}{\alpha} v_s$,  the flow direction of the HL is defined as $x$-direction.  After adopting the approximation in the previous section,  the analytical expression of 
the correction $\Delta v_{d}$ of  the helical magnetic state is
\begin{equation}
	\Delta v_{d} \approx-\chi_d^{HL} (v_{d0})^{\frac{d-2}{2}} \label{Eq:helix}
\end{equation}
where $
	\chi_d^{HL} =
		\frac{(K_xK_y)^{-1/2}|\rho_1|^2\Lambda_0g_0^3}{4\alpha D}, \text{for } d=2 $ $
		\frac{(K_xK_yK_z)^{-1/2}|\rho_1|^2\Lambda_0g_0^{7/2}}{2\sqrt{2}\pi(\alpha D)^{1/2}}, \text{for } d=3 $ with $g_0=\frac{\pi}{a}$.
Despite different magnetic state nature, the $\Delta v_{d}$ as a function $v_{d0}$ in Eq.~\eqref{Eq:helix} for the HL displays a consistent scaling behavior as the one of SkX shown in Eq.~\eqref{Eq:final}.

\emph{Numerical evaluation.---} To further justify our analytical results,  we calculate the $\Delta v_d$ numerically  according to Eqs.~\eqref{Eq:velocity} and \eqref{Eq:helical}. For simplicity, we set the elastic coefficient $K_j$ as isotropic with $K\equiv K_j$. Figs.~\ref{fig:fig2}(a) and (b) display the correction $\Delta v_d$ as a function of $v_{d0}$ of HL. Note that the zero-drift velocity limit should be ignored since our theoretical consideration is for the flow region, where the drift velocity is far from zero.  In the large $v_d$ or the flow limit, where the pinning effect can be treated as a perturbation,  the plots clearly indicate $\Delta v_d \propto (v_{d0})^{\frac{d-2}{2}}$. The square root behavior in 3D ($d=3$) is explicitly checked with the log-log plot (inset of Fig.~\ref{fig:fig2}(b)). 

Figs.~\ref{fig:fig2}(c) shows that the ratio between the transverse and longitudinal component is approaching $G/\alpha D$ in  SkX case, being consistent with Eq.~\eqref{Eq:final}. This gyrotropic type correction is inherited from the Magnus forces in the Thiele equation, and this correction also implies that there exists a net change on the skyrmion Hall angle due to the impurities. 
Moreover, the angular dependence of the total correction $|\Delta \bm{v}_d|$  is shown in Figs.~\ref{fig:fig2}(e), where the anisotropy is very small.  The $\Delta v_d$ as a function $v_{d0}$ also displays distinct scaling behavior between 2D [Figs.~\ref{fig:fig2}(c)] and 3D case [Fig.~\ref{fig:fig2}(f)].  

Overall, the scaling behavior of SkX is similar to that of the HL.  Moreover,  the intrinsic drift velocity $v_{d0}$ is linearly proportional to the current density $j_s$ for both SkX and HL ($v_{d0} \propto j_s$) at large $j_s$. As a result,    we can replace $v_{d0} $ with $j_s$  in the scaling relation, i.e., $\Delta v_d\propto (j_s)^{\frac{d-2}{2}}$. It is worth noting that the charge density wave also respects this scaling relation \cite{Sneddon1982}, despite its distinct microscopic nature. 


\emph{Physical interpretation.---}Now we provide a physical interpretation of the observed scaling behavior: $\Delta v_d \propto (v_{d0})^{\frac{d-2}{2}}$.  As we mentioned earlier, the dominant contribution to the drift velocity correction arises from the excitation of elastic modes. Hence, we expect the correction to be proportional to the number of excited elastic modes at a fixed $v_{d0}$.  For the SkX case, these modes follow the dispersion: $v_{d0}|\bm{g}_{j}|= \mathcal{D}(\bm{k})/\sqrt{G^2+\alpha^2D^2}$, which can be rewritten as $v_{d0}=\bm{k}^d/2m'$ with $m'=2\pi\sqrt{G^2+\alpha^2D^2}/(\sqrt{3}aK)$. Next, the problem is mapped to evaluate the density of states of free fermions with an energy $v_{d0}$. Recall that the density of states of free fermion $N(E) \propto E^{\frac{d-2}{2}}$ at energy $E$. Hence, it is expected that the correction follows the same scaling:  $\Delta v_d \propto (v_{d0})^{\frac{d-2}{2}}$ according to this argument.  We emphasize that the microscopic nature of the density waves in this argument is not essential, which mainly stems from the long-wave characteristic of elastic modes. This explains why the HL and charge density wave also follow the same scaling behavior.

\emph{Micromagnetic simulation.---} We now further validate our theory through solving the Landau–Lifshitz–Gilbert (LLG) equation with the spin transfer torque effect \cite{vansteenkiste_design_2014,slonczewski_current_driven_1996, berger_emission_1996, zhang_roles_2004,koshibae_theory_2018} (for details see SM).
\begin{figure}
	\centering
	\includegraphics[width=1\linewidth]{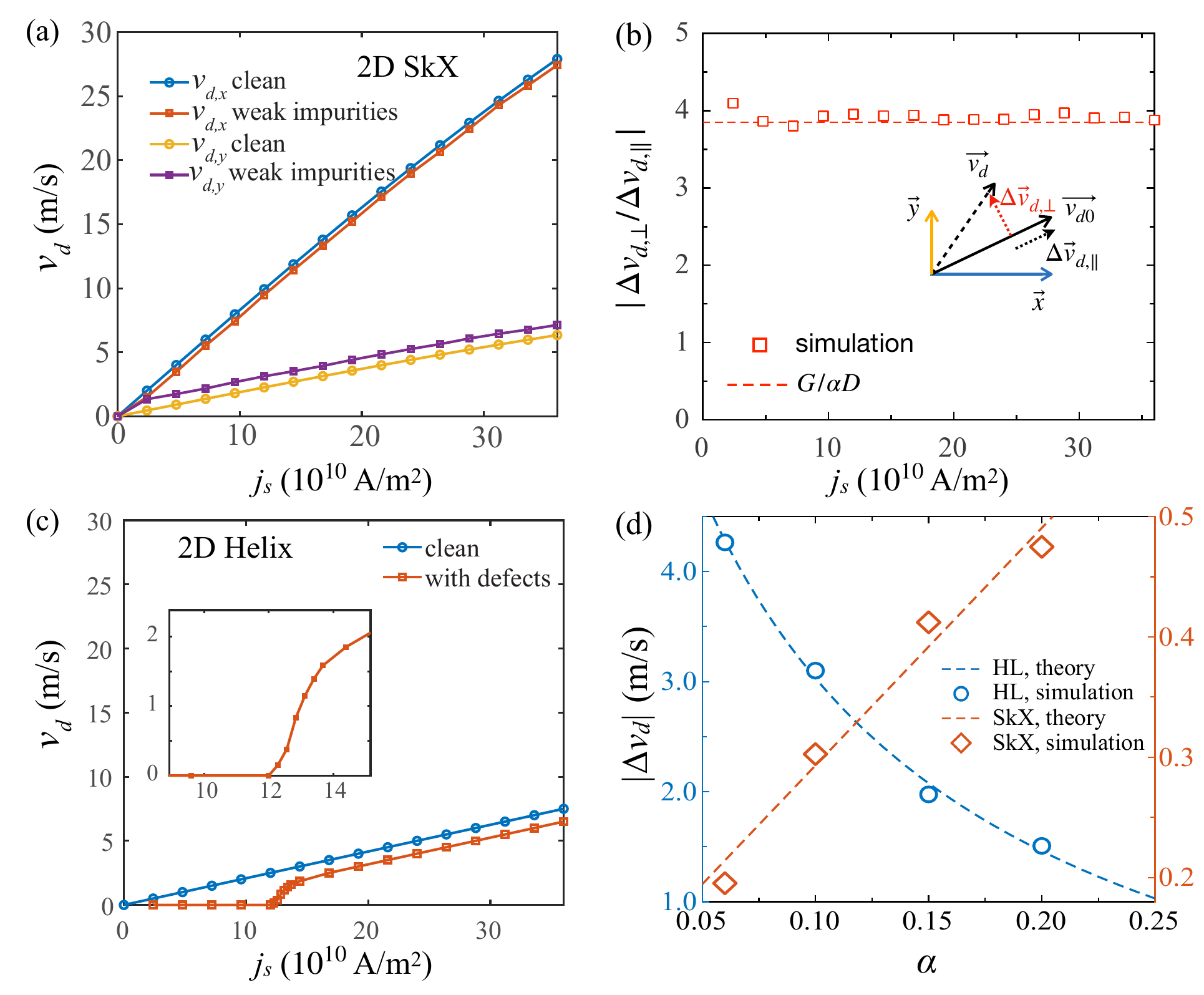}
 	\caption{Simulation results of LLG equation.  (a) and (c) The drift velocity $v_d$ versus current density $j_s$ (in units of $10^{10}$ A/m$^2$) of the 2D SkX and HL at the clean and impurity case. (b) The ratio between the transverse ($\Delta v_{d,\perp}$) and the longitudinal ($\Delta v_{d,\parallel}$) drift velocity correction versus the current density deduced from (a) for 2D SkX, where the $G/\alpha D$ ratio is highlighted (red dashed line). The coordinate relation between different vectors are shown in the inset.  (d) the drift velocity correction as a function of the damping parameter $\alpha$ at $j_s=2\times 10^{11} \text{ A/m}^2$ (only $v_{d,x}$ is used for the SkX). The micromagnetic simulation is in agreement with the theoretical expectation  (dashed lines).  For (a) to (c), $\alpha=0.2$ and $\beta=0.5\alpha$ are employed in the simulations.} 
  \label{fig:fig3}
\end{figure}
The calculated drift velocity $v_d$ versus current density $j_s$ curves are shown in Fig.~\ref{fig:fig3} for both the SkX and HL. 
For simplicity, we mainly focus on 2D SkX and HL with weak impurities here, where our analytical expressions from perturbation theory are applicable. 

Figs.~\ref{fig:fig3}(a) and (c)  show  $v_d$ in the clean and disordered case with $\alpha=0.2$. The correction between these two cases at both SkX and HL is indeed insensitive to the current density within the flow limit. It is noteworthy that due to the gyrodynamics,  the SkX exhibits a much smaller depinning critical current density.   
Fig.~\ref{fig:fig3}(b) is to show that the correction along the transverse direction is obviously larger than the longitudinal one with the ratio $\sim  G/\alpha D $, being consistent with Fig.~\ref{fig:fig2}(c). 
Interestingly, the longitudinal correction versus the damping parameter $\alpha$ of 2D SkX and HL show a positive and negative correlation, respectively [Fig.~\ref{fig:fig3}(d)], which is also consistent with results of our developed model [see Fig.~\ref{fig:fig2}(d)]. It can also be seen that the impurity correction along the longitudinal direction is typically much stronger in  HL than in   SkX as $\chi^{SkX}_{d}/\chi^{HL}_{d} \sim \alpha^2$.  These distinct features between SkX and HL highlight the importance of the nontrivial topology in the sliding dynamics of density waves.  



%

\emph{Discussions.---} We have provided a thorough analysis of the sliding dynamics exhibited by the SkX and HL phases, highlighting both their similarities and differences in terms of density waves sliding with distinct topologies.  Our theory could have broader applications.  For instance, one can explore the relationship between the topological Hall effect and the current density  in the flow region.   In the clean limit, the universal linear current-velocity relation $v_{d0}\propto j_s$ implies that the topological Hall resistivity $\rho_{xy}$, proportional to $|(\bm{v}_s-\bm{v}_{d0})\times \bm{B}^{e}|/|\bm{v}_s|$ \cite{Schulz2012}, is expected to exhibit a plateau in the flow region, as illustrated in Fig.~\ref{fig:fig1}(c). In the presence of impurities, the topological Hall resistivity is modified to $\rho_{xy}\propto |(\bm{v}_s-\bm{v}_{d0}-\Delta \bm{v}_d)\times \bm{B}^{e}|/|\bm{v}_s|$. Considering that $\Delta v_{d}\propto (v_{d0})^{(d-2)/2}$, we anticipate a modified relation $\rho_{xy}=a +bj_s^{-2+d/2}$, where $a$ and $b$ remain independent of the current magnitude $j_s$. The second term, $bj_s^{-2+d/2}$, represents the correction from impurities. Consequently, we expect that the $\rho_{xy}$-$j_s$ plateau in the flow region will gradually diminish with increasing disorder. In addition to the topological Hall measurements, the noise features of moving SkX and HL can also be distinct and might be employed to distinguish the SkX and HL phases \cite{Reichhardt_2016,  Reichhardt2017_noise, sato_slow_2019, Reichhardt_RMP}

It is worth noting that the Thiele equation approach, as a particle-based approach, is an approximated method. Despite its simplification, it has been proven to be a powerful tool in describing the dynamics of spin textures in many real experiments~\cite{buttner_dynamics_2015, jiang_direct_2017, litzius_skyrmion_2017, wang_thermal_2020}. In our developed model, the chiral magnets display crystal flowing and the deformation within each SkX is tiny (see SM Sec.~VII).  As a result,   the approximation for the Thiele method is justified in our case.  Consistently, we also have used the micromagnet simulations to verify the key predictions from the Thiele equation approach.  Moreover, the particle-based method itself has wide applications in the pinning dynamics of many other systems \cite{FISHER1998113, Reichhardtreview2017, Reichhardt_RMP}, such as superconducting vortices \cite{Jensen1988, Higgins1993,Reichhardt1997, Nori1998}, Wigner crystal \cite{Millis1994, Reichhardt2022, Madathil2023}.  Hence, the developed theory in this work could exhibit wide implications.


\emph{Acknowledgment.---}	We thank Max Birch and Yoshinori Tokura for presenting us with their Hall measurement data on the SkX, which motivated this study. We also thank Wataru Koshibae for useful discussions. N.N. was supported by JSTCREST Grants No.JMPJCR1874. Y.M.X. and Y.L. acknowledge financial support from the RIKEN Special Postdoctoral Researcher(SPDR) Program.

 \clearpage
	
	\onecolumngrid
	\begin{center}
		\textbf{\large Supplementary Material for  \lq\lq{} Sliding Dynamics of Current-Driven Skyrmion Crystal and Helix in Chiral Magnets \rq\rq{}}\\[.2cm]
		Ying-Ming Xie,$^{1}$ Yizhou Liu,$^{1}$ Nato Nagaosa,$^{1}$ \\[.1cm]
		{\itshape ${}^1$    RIKEN Center for Emergent Matter Science (CEMS), Wako, Saitama 351-0198, Japan}\\[1cm]
	\end{center}
	\setcounter{equation}{0}
	\setcounter{section}{0}
	\setcounter{figure}{0}
	\setcounter{table}{0}
	\setcounter{page}{1}
	\renewcommand{\theequation}{S\arabic{equation}}
	\renewcommand{\thetable}{S\arabic{table}}
	\renewcommand{\thesection}{\Roman{section}}
	\renewcommand{\thefigure}{S\arabic{figure}}
	\renewcommand{\bibnumfmt}[1]{[S#1]}
	\renewcommand{\citenumfont}[1]{#1}
	\makeatletter
	
	\onecolumngrid
	
	\maketitle
	\section{The sliding dynamics of Skyrmion Crystal  with pining and deformation effects}

\subsection{The skyrmion dynamics and Thiele equation}
 From the Landau-Lifshitz-Gilbert equation, it was obtained that	the current-driven  skyrmion  dynamics are captured by the Thiele equation:
 	\begin{equation}
 		\bm{G}\times (\bm{v}_s-\bm{v}_d)+\bm{D}(\beta \bm{v}_s-\alpha \bm{v}_d)+\bm{F}=0.
 	\end{equation}
One can rewrite the equation as
\begin{eqnarray}
		-G(v_{sy}-v_{dy})+D(\beta v_{sx}-\alpha v_{dx})+F_x=0,\\
	G(v_{sx}-v_{dx})+D(\beta v_{sy}-\alpha v_{dy})+F_y=0.
\end{eqnarray}
In the matrix form:
\begin{equation}
	\begin{pmatrix}
		G& \alpha D\\
		\alpha D&-G
	\end{pmatrix}\begin{pmatrix}
	v_{dx}\\ v_{dy}
\end{pmatrix}=\begin{pmatrix}
G&\beta D\\\beta D& -G
\end{pmatrix}\begin{pmatrix}
v_{sx}\\v_{sy}
\end{pmatrix}+\begin{pmatrix}
F_y\\F_x
\end{pmatrix}.
\end{equation}
Then,
\begin{equation}
	\begin{pmatrix}
		v_{dx}\\ v_{dy}
	\end{pmatrix}=	\begin{pmatrix}
	G& \alpha D\\
	\alpha D&-G
\end{pmatrix}^{-1}\begin{pmatrix}
		G&\beta D\\\beta D& -G
	\end{pmatrix}\begin{pmatrix}
		v_{sx}\\v_{sy}
	\end{pmatrix}+	\begin{pmatrix}
	G& \alpha D\\
	\alpha D&-G
\end{pmatrix}^{-1}\begin{pmatrix}
		F_y\\F_x
	\end{pmatrix}.
\end{equation}

\begin{eqnarray}
	\begin{pmatrix}
		v_{dx}\\ v_{dy}
	\end{pmatrix}&&=	\frac{1}{G^2+\alpha^2D^2}\left[\begin{pmatrix}
	G^2+\alpha\beta D^2&GD(\beta-\alpha)\\GD(\alpha-\beta)&G^2+\alpha\beta D^2
\end{pmatrix}\begin{pmatrix}
		v_{sx}\\v_{sy}
	\end{pmatrix}+	\begin{pmatrix}
	G&\alpha D\\\alpha D&-G
\end{pmatrix}\begin{pmatrix}
0&1\\1&0
\end{pmatrix}\begin{pmatrix}
		F_x\\F_y
	\end{pmatrix}\right],\nonumber\\
&&=\frac{1}{G^2+\alpha^2D^2}\left[\begin{pmatrix}
	G^2+\alpha\beta D^2&GD(\beta-\alpha)\\GD(\alpha-\beta)&G^2+\alpha\beta D^2
\end{pmatrix}\begin{pmatrix}
	v_{sx}\\v_{sy}
\end{pmatrix}+	\begin{pmatrix}
\alpha D&G\\
-G&\alpha D
\end{pmatrix}\begin{pmatrix}
	F_x\\F_y
\end{pmatrix}\right]\label{Eq:thiele2}
\end{eqnarray}
Without loss of generality, we can choose the current direction to be $x$-direction: $\bm{v_s}=(v_{s},0)$. When the pinning force is set to be $F=0$, one can solve 
\begin{equation}
	\bm{v_{d\parallel,0}}= \frac{G^2+\alpha\beta D^2}{G^2+\alpha^2D^2} \bm{v_s}, 	\bm{v_{d\perp,0}}=\frac{(\alpha-\beta)GD}{G^2+\alpha^2D^2}\hat{z}\times \bm{v_s}.
\end{equation}
 Therefore, the longitudinal drift velocity $v_{dx}$ is proportional to the electric current when the force $\bm{F}$ is neglectable.   In a general direction, we can  write the  intrinsic drift velocity as
\begin{eqnarray}
\bm{v}_{d0}&&=\frac{1}{G^2+\alpha^2D^2}\begin{pmatrix}
		G^2+\alpha\beta D^2&GD(\beta-\alpha)\\GD(\alpha-\beta)&G^2+\alpha\beta D^2
\end{pmatrix}\begin{pmatrix}
		v_{sx}\\v_{sy}	\end{pmatrix}\\
&&=\sqrt{\frac{1+\beta^2\gamma^2}{1+\alpha^2\gamma^2}}\begin{pmatrix}
	\cos\theta_{SkH}&-\sin\theta_{SkH}\\
 \sin\theta_{SkH}& \cos\theta_{SkH}
\end{pmatrix}\begin{pmatrix}
		v_s\cos\theta_s\\
  v_s\sin\theta_s
\end{pmatrix}\\
&&=v_{d0}\begin{pmatrix}
\cos(\theta_s+\theta_{SkH})\\\sin(\theta_s+\theta_{SkH})
		\end{pmatrix}.
\end{eqnarray}
Here, the angle $\theta_s$ is to characterize the applied current direction, the skyrmion Hall angle $\theta_{SkH}=\text{atan}\frac{\gamma(\alpha-\beta)}{1+\alpha\beta\gamma^2}$ with $\gamma=\frac{D}{G}$, and the magnitude of drift velocity $v_{d0}=|\bm{v}_{d0}|=v_s\sqrt{\frac{1+\beta^2\gamma^2}{1+\alpha^2\gamma^2}}$.

Now we show that the Thiele equation respects rotational symmetry with the principal axis along $z$-direction from Eq.~\eqref{Eq:thiele2}. The rotational operator is defined as $R_z=\begin{pmatrix}
	\cos\phi&-\sin(\phi)\\
	 \sin(\phi)&\cos(\phi)
\end{pmatrix}$ with $\phi$ as the rotational angle. Under this rotational operation, Eq.~\eqref{Eq:thiele2} becomes
\begin{eqnarray}
	R_z\begin{pmatrix}
		v_{dx}\\ v_{dy}
	\end{pmatrix}&&=	\frac{1}{G^2+\alpha^2D^2}\left[\begin{pmatrix}
		G^2+\alpha\beta D^2&GD(\beta-\alpha)\\GD(\alpha-\beta)&G^2+\alpha\beta D^2
	\end{pmatrix}\begin{pmatrix}
		v_{sx}\\v_{sy}
	\end{pmatrix}+	\begin{pmatrix}
		G&\alpha D\\\alpha D&-G
	\end{pmatrix}\begin{pmatrix}
		0&1\\1&0
	\end{pmatrix}\begin{pmatrix}
		F_x\\F_y
	\end{pmatrix}\right],\nonumber\\
	&&=\frac{1}{G^2+\alpha^2D^2}\left[ R_z\begin{pmatrix}
		G^2+\alpha\beta D^2&GD(\beta-\alpha)\\GD(\alpha-\beta)&G^2+\alpha\beta D^2
	\end{pmatrix}R_z^{-1}R_z\begin{pmatrix}
		v_{sx}\\v_{sy}
	\end{pmatrix}+	R_z\begin{pmatrix}
		\alpha D&G\\
		-G&\alpha D
	\end{pmatrix}R_z^{-1}R_z\begin{pmatrix}
		F_x\\F_y
	\end{pmatrix}\right]\label{Eq:thiele}
\end{eqnarray}
It is easy to show 
\begin{equation}
	R_z\begin{pmatrix}
		A&B\\-B&A
	\end{pmatrix}R_z^{-1}=\begin{pmatrix}
	A&B\\-B&A
	\end{pmatrix}
\end{equation}
with $A$ and $B$ as  constant.  The Eq.~\eqref{Eq:thiele} is simplified as 
\begin{eqnarray}
	R_z\begin{pmatrix}
		v_{dx}\\ v_{dy}
	\end{pmatrix}&&=	\frac{1}{G^2+\alpha^2D^2}\left[\begin{pmatrix}
		G^2+\alpha\beta D^2&GD(\beta-\alpha)\\GD(\alpha-\beta)&G^2+\alpha\beta D^2
	\end{pmatrix}\begin{pmatrix}
		v_{sx}\\v_{sy}
	\end{pmatrix}+	\begin{pmatrix}
		G&\alpha D\\\alpha D&-G
	\end{pmatrix}\begin{pmatrix}
		0&1\\1&0
	\end{pmatrix}\begin{pmatrix}
		F_x\\F_y
	\end{pmatrix}\right],\nonumber\\
	&&=\frac{1}{G^2+\alpha^2D^2}\left[ \begin{pmatrix}
		G^2+\alpha\beta D^2&GD(\beta-\alpha)\\GD(\alpha-\beta)&G^2+\alpha\beta D^2
	\end{pmatrix}R_z\begin{pmatrix}
		v_{sx}\\v_{sy}
	\end{pmatrix}+	\begin{pmatrix}
		\alpha D&G\\
		-G&\alpha D
	\end{pmatrix}R_z\begin{pmatrix}
		F_x\\F_y
	\end{pmatrix}\right]\label{Eq:thiele}
\end{eqnarray}

Hence, we have shown that the Thiele equation respects out-of-plane rotational symmetry.
\subsection{The correction on the drifted velocity due to the pining and deformation effects}
Let us define the displacement of skyrmion lattice as $\bm{u}(\bm{r},t)$ so that the drift velocity  $\bm{v_d}(\bm{r},t)=\frac{\partial \bm{u}(\bm{r},t)}{\partial t}$.
The force is given by
\begin{eqnarray}
	\bm{F}(\bm{r},t)&&=\bm{F}_{imp}+\bm{F}_{de}\\
\bm{F}_{imp}&&=	-\sum_{i}\nabla U(\bm{r}+\bm{u}(\bm{r},t)-\bm{r}_i)\rho(\bm{r})=\bm{f}_{imp}(\bm{r}+\bm{u}(\bm{r},t))\rho(\bm{r})\\
\bm{F}_{de}&&=	\int d\bm{r'} \mathcal{D}(\bm{r}-\bm{r'})\bm{u}(\bm{r'},t'),
\end{eqnarray}
where $\bm{F}_{im}$ describes the pining effect from impurities and  $\bm{F}_{de}$ arises from the deformation of skyrmion lattice, $U(\bm{r}-\bm{r}_i)$ is the impurity potential around the site $\bm{r}_i$. $\rho(\bm{r})$ is the skyrmion densities.
 
 \begin{eqnarray}
\frac{\partial \bm{u}(\bm{r},t)}{\partial t}&&=	\frac{1}{G^2+\alpha^2D^2}\left[\begin{pmatrix}
 		G^2+\alpha\beta D^2&GD(\beta-\alpha)\\GD(\alpha-\beta)&G^2+\alpha\beta D^2
 	\end{pmatrix}\begin{pmatrix}
 		v_{sx}\\v_{sy}
 	\end{pmatrix}+
 \begin{pmatrix}
 	\alpha D&G\\
 	-G&\alpha D
 \end{pmatrix}	\int d\bm{r'} \mathcal{D}(\bm{r}-\bm{r'})\begin{pmatrix}
 		u_x(\bm{r'},t)\\	u_y(\bm{r'},t)
 	\end{pmatrix}\right]\nonumber\\
&&+\frac{1}{G^2+\alpha^2D^2}\begin{pmatrix}
	\alpha D&G\\
	-G&\alpha D
\end{pmatrix}\bm{f}_{imp}(\bm{r}+\bm{u}(\bm{r},t))\rho(\bm{r}) .\label{Eq:motion}
 \end{eqnarray}

 The displacement vector can be expanded  around the uniform motion,
\begin{equation}
	\bm{u}(\bm{r},t)=\bm{v_d}t+\tilde{\bm{u}}(\bm{r},t).
\end{equation} 
Here, $\bm{v_d}$ is the dominant uniform skyrmion motion velocity,  $\tilde{\bm{u}}(\bm{r},t)$ characterizes a small non-uniform motion. Then  the equation of motion is written as

\begin{eqnarray}
	\left[\frac{\partial }{\partial t}-\frac{1}{G^2+\alpha^2D^2}	\begin{pmatrix}
		\alpha D&G\\
		-G&\alpha D
	\end{pmatrix}\int d\bm{r'} \mathcal{D}(\bm{r}-\bm{r'})\right]\begin{pmatrix}
	\bm{\tilde{u}}_x(\bm{r'},t)\\	\bm{\tilde{u}}_y(\bm{r'},t)
\end{pmatrix}&&=\frac{1}{G^2+\alpha^2D^2}\begin{pmatrix}
G^2+\alpha\beta D^2&GD(\beta-\alpha)\\GD(\alpha-\beta)&G^2+\alpha\beta D^2
\end{pmatrix}\begin{pmatrix}
v_{sx}\\v_{sy}
\end{pmatrix}-\bm{v_d}\nonumber\\&&+\frac{1}{G^2+\alpha^2D^2}\begin{pmatrix}
\alpha D&G\\
-G&\alpha D
\end{pmatrix}\bm{f}_{imp}(\bm{r}+\bm{v_d}t+\tilde{\bm{u}}(\bm{r},t))\rho(\bm{r}).\nonumber\\
\label{Eq:motion2}
\end{eqnarray}
Here, we have used $\mathcal{D}(\bm{q})=0$  in the long wave limit  ($\bm{q}\rightarrow0$) so that $\int  d\bm{r'}\mathcal{D}(\bm{r}-\bm{r'})=0$.

Let us try to solve the Green's function of  the operator at the left-hand side, which is given by

	\begin{eqnarray}
		\left[\frac{\partial }{\partial t}-\frac{1}{G^2+\alpha^2D^2}	\begin{pmatrix}
			\alpha D&G\\
			-G&\alpha D
		\end{pmatrix}\int d\bm{r'} \mathcal{D}(\bm{r}-\bm{r'})\right]\mathcal{G}(\bm{r}',t)=\delta(t)\delta(\bm{r})\begin{pmatrix}
		1&0\\0&1
	\end{pmatrix}
\end{eqnarray}
It is more economical to work in the momentum space with
\begin{equation}
	\mathcal{G}(\bm{r},t)=\int \frac{d\omega}{2\pi}\int \frac{d^d \bm{k}}{(2\pi)^d}e^{-i\omega t+i\bm{k}\cdot\bm{r}}\mathcal{G}(\bm{k},\omega).
\end{equation}
Let us define $\mathcal{D}(\bm{r})=\int \frac{d^d q}{(2\pi)^d} e^{i\bm{q}\cdot \bm{r}} \mathcal{D}(\bm{q})$, and then
\begin{eqnarray}
	\int d\bm{r'} \int \mathcal{D}(\bm{r}-\bm{r'})\mathcal{G}(\bm{r'},t)&&=\int d\bm{r'} \int \frac{d^d \bm{q}}{(2\pi)^d} e^{i\bm{q}\cdot (\bm{r}-\bm{r'})} \mathcal{D}(\bm{q}) \int \frac{d^d \bm{k}}{(2\pi)^d} \int \frac{d\omega}{2\pi} \mathcal{G}(\bm{k},\omega )e^{i(\bm{k}\cdot\bm{r'}-\omega t)}\nonumber\\
	&&= \int \frac{d^{d}\bm{k}}{(2\pi)^d}\int \frac{d\omega}{(2\pi)}\mathcal{G}(\bm{k},\omega) \mathcal{D}(\bm{k}) e^{i(\bm{k}\cdot \bm{r}-\omega t)}.
\end{eqnarray}
In the momentum space, we  find
\begin{equation}
	\mathcal{G}^{-1}(\bm{k},\omega)=-i\omega-\frac{1}{G^2+(D\alpha)^2}\begin{pmatrix}
		\alpha D&G\\
		-G&\alpha D
	\end{pmatrix}  \mathcal{D}(\bm{k})
\end{equation}
Therefore, Eq.~(\ref{Eq:motion2}) can be rewritten as
\begin{align}
	\bm{\tilde{u}}(\bm{r},t)&=\int d\bm{r'}\int dt' \mathcal{G}(\bm{r}-\bm{r'},t-t')\{\frac{1}{G^2+\alpha^2D^2}\begin{pmatrix}
		G^2+\alpha\beta D^2&GD(\beta-\alpha)\\GD(\alpha-\beta)&G^2+\alpha\beta D^2
	\end{pmatrix}\begin{pmatrix}
		v_{sx}\\v_{sy}
	\end{pmatrix}-\bm{v_d}\\
&\nonumber+\frac{1}{G^2+\alpha^2D^2}\begin{pmatrix}
	\alpha D&G\\
	-G&\alpha D
\end{pmatrix}\bm{f}_{imp}(\bm{r'}+\bm{v_d}t'+\bm{\tilde{u}}(\bm{r'},t'))\rho(\bm{r'})\}.
\end{align}

In the flow limit, the perturbation from the deformation and impurity can be regarded as small in comparison with the leading order term. As a result, the displacement vector can be expanded as
\begin{eqnarray}
		\tilde{\bm{u}}_0(\bm{r},t)=&&\int d\bm{r'}\int dt' \mathcal{G}(\bm{r}-\bm{r'},t-t')\left[\frac{1}{G^2+\alpha^2D^2}\begin{pmatrix}
		G^2+\alpha\beta D^2&GD(\beta-\alpha)\\GD(\alpha-\beta)&G^2+\alpha\beta D^2
	\end{pmatrix}\begin{pmatrix}
		v_{sx}\\v_{sy}
	\end{pmatrix} -\bm{v_d}\right],\\
	\tilde{\bm{u}}_1(\bm{r},t)=&&\frac{1}{G^2+\alpha^2D^2} \int d\bm{r'}\int d t' \mathcal{G}(\bm{r}-\bm{r'},t-t')\begin{pmatrix}
		\alpha D&G\\
		-G&\alpha D
	\end{pmatrix} \bm{f_{imp}}(\bm{r'}+\bm{v_d}t')\rho(\bm{r'}),\\
\tilde{\bm{u}}_2(\bm{r},t)=&&\frac{1}{G^2+\alpha^2D^2} \int d\bm{r'}\int d t' \mathcal{G}(\bm{r}-\bm{r'},t-t')\begin{pmatrix}
	\alpha D&G\\
	-G&\alpha D
\end{pmatrix} \nabla\bm{f_{imp}}(\bm{r'}+\bm{v_d}t')\cdot \bm{\tilde{u}_1}(\bm{r'},t')\rho(\bm{r'}).
\end{eqnarray}
Here, $\bm{u}_0$, $\bm{u}_1$, and  $\bm{u}_2$ are the leading, first, and second order terms, respectively.
Next, let us evaluate 
 the volume-average velocity
 \begin{equation}
 	\left<\frac{\partial \tilde{\bm{u}}(\bm{r},t)}{\partial t}\right>= 	\left<\frac{\partial \tilde{\bm{u}}_0(\bm{r},t)}{\partial t}\right>+\left<\frac{\partial \tilde{\bm{u}}_2(\bm{r},t)}{\partial t}\right>
 \end{equation}
 Note the fact that under the impurity average
 $\overline{f_{imp}(\bm{r'}+vt')}=0$ has been used so that $\bm{u}_1(\bm{r},t)$ would not contribute directly. Since non-uniform motion must vanish over the volume average, we can obtain a self-consistent equation for the velocity $\bm{v_d}$. Next, let us work out the self-consistent equation for  $\bm{v_d}$.

 The leading order
\begin{eqnarray}
	&&\frac{\partial \bm{\tilde{u}}_0(\bm{r},t)}{\partial t}=\int d\bm{r'}\int dt' \frac{\mathcal{G}(\bm{r}-\bm{r'},t-t')}{\partial t}\left[\frac{1}{G^2+\alpha^2D^2}\begin{pmatrix}
		G^2+\alpha\beta D^2&GD(\beta-\alpha)\\GD(\alpha-\beta)&G^2+\alpha\beta D^2
	\end{pmatrix}\begin{pmatrix}
	v_{sx}\\v_{sy}
\end{pmatrix}-\bm{v_d}\right]\nonumber\\
&&= \int d\bm{r'}\int dt' \int \frac{d^d \bm{k}}{(2\pi)^d} \int \frac{d\omega}{2\pi} e^{i\bm{k}\cdot(\bm{r}-\bm{r'})-i\omega (t-t')} (-i\omega)\mathcal{G}(\bm{k},\omega)\left[\frac{1}{G^2+\alpha^2D^2}\begin{pmatrix}
	G^2+\alpha\beta D^2&GD(\beta-\alpha)\\GD(\alpha-\beta)&G^2+\alpha\beta D^2
\end{pmatrix}\begin{pmatrix}
v_{sx}\\v_{sy}
\end{pmatrix}-\bm{v_d}\right]\nonumber\\
&&=\lim_{\omega \rightarrow 0}-i\omega \mathcal{G}(\bm{k}=0,\omega)\left[\frac{1}{G^2+\alpha^2D^2}\begin{pmatrix}
	G^2+\alpha\beta D^2&GD(\beta-\alpha)\\GD(\alpha-\beta)&G^2+\alpha\beta D^2
\end{pmatrix}\begin{pmatrix}
v_{sx}\\v_{sy}
\end{pmatrix}-\bm{v_d}\right]\nonumber\\
&&=\left[\frac{1}{G^2+\alpha^2D^2}\begin{pmatrix}
	G^2+\alpha\beta D^2&GD(\beta-\alpha)\\GD(\alpha-\beta)&G^2+\alpha\beta D^2
\end{pmatrix}\begin{pmatrix}
	v_{sx}\\v_{sy}
\end{pmatrix}-\bm{v_d}\right]
\end{eqnarray}

As mentioned the $\left<\frac{\partial \tilde{u}_1(\bm{r},t)}{\partial t}\right>$ would not contribute, now let us show it explicitly. Recall that 
\begin{equation}
	\tilde{\bm{u}}_1(\bm{r},t)=\frac{1}{G^2+\alpha^2D^2}\int d\bm{r'}\int dt' \mathcal{G}(\bm{r}-\bm{r'},t-t')\begin{pmatrix}
		\alpha D&G\\
		-G&\alpha D
	\end{pmatrix} \bm{f_{imp}}(\bm{r'}+\bm{v_d}t')\rho(\bm{r'})
\end{equation}
Then,
\begin{eqnarray}
 	\frac{\partial\bm{\tilde{u}}_1(\bm{r},t)}{\partial t}&&=\int d\bm{r'}\int dt' \frac{1}{G^2+\alpha^2D^2}  \frac{\mathcal{G}(\bm{r}-\bm{r'},t-t')}{\partial t}\begin{pmatrix}
 		\alpha D&G\\
 		-G&\alpha D
 	\end{pmatrix} \bm{f_{imp}}(\bm{r'}+\bm{v_d}t')\rho(\bm{r'})\\
&&=- \frac{1}{G^2+\alpha^2D^2} \int d\bm{r'}\int dt' \int \frac{d^d \bm{k}}{(2\pi)^d} \int \frac{d\omega}{2\pi} e^{i\bm{k}\cdot(\bm{r}-\bm{r'})-i\omega (t-t')} (-i\omega) \mathcal{G}(\bm{k},\omega)\begin{pmatrix}
	\alpha D&G\\
	-G&\alpha D
\end{pmatrix}\times\\
&&\int \frac{d^{d}\bm{q}}{(2\pi)^d}\begin{pmatrix}
i q_x \\ iq_y
\end{pmatrix}\overline{U_{\bm{q}}}e^{i\bm{q}\cdot(\bm{r'}+\bm{v_d}t')}\rho(\bm{r}')\\
&&=0,
\end{eqnarray}
because after averaging over the impurity  configurations, $\overline{U_{\bm{k}}}=0$.

Now let us look at the second-order term

\begin{eqnarray}
	\frac{\partial \bm{\tilde{u}}_2(\bm{r},t)}{\partial t}=&&\frac{1}{G^2+\alpha^2D^2} \int d\bm{r'}\int d t' \frac{ \partial \mathcal{G}(\bm{r}-\bm{r'},t-t')}{\partial t}\begin{pmatrix}
		\alpha D&G\\
		-G&\alpha D
	\end{pmatrix} \nabla\bm{f_{imp}}(\bm{r'}+\bm{v_d}t')\cdot \bm{\tilde{u}_1}(\bm{r'},t')\rho(\bm{r'}).
\end{eqnarray}
Note that 
	\begin{eqnarray}
		\nabla\bm{f_{imp}}(\bm{r'}+\bm{v_d}t')\cdot \bm{\tilde{u}_1}(\bm{r'},t')&&=-\int\frac{d^{d} \bm{q}}{(2\pi)^d} U(\bm{q})e^{i\bm{q}\cdot(\bm{r'}+\bm{v_d}t')}
		\begin{pmatrix} 
q_{x}^2&q_{x}q_{y}\\q_{x}q_{y}&	q_{y}^2
		\end{pmatrix}	\begin{pmatrix}
			\tilde{u}_{1x}(\bm{r'},t')\\	\tilde{u}_{1y}(\bm{r'},t')
		\end{pmatrix}
	\end{eqnarray}
Substitute the form of $\tilde{\bm{u}}_1(\bm{r'},t)$,

\begin{eqnarray}
	\frac{\partial \bm{\tilde{u}}_2(\bm{r},t)}{\partial t}=&&-\frac{1}{(G^2+\alpha^2D^2)^2} \int d\bm{r'}\int d t' \frac{ \partial \mathcal{G}(\bm{r}-\bm{r'},t-t')}{\partial t}\begin{pmatrix}
		\alpha D&G\\
		-G&\alpha D
	\end{pmatrix} \rho(\bm{r'})\times \nonumber\\
&&\int\frac{d^{d} \bm{q}_1}{(2\pi)^d} U(\bm{q}_1)e^{i\bm{q}_1(\cdot\bm{r'}+\bm{v_d}t')}
	\begin{pmatrix} 
q_{1x}^2&q_{1x}q_{1y}\\q_{1x}q_{1y}&	q_{1y}^2
		\end{pmatrix}\times\nonumber\\
&& \int dt''\int d\bm{r''} \mathcal{G}(\bm{r}'-\bm{r''},t'-t'')\begin{pmatrix}
	\alpha D&G\\
	-G&\alpha D
\end{pmatrix}\rho(\bm{r''})\int \frac{d^d \bm{q}_2}{(2\pi)^d}\begin{pmatrix}
i q_{2x}\\ iq_{2y}
\end{pmatrix} U(\bm{q}_2)e^{i\bm{q}_2\cdot(\bm{r''}+\bm{v_d}t'')}	
\end{eqnarray}
Then write the terms at the right-hand side of the equation with their Fourier components,
\begin{eqnarray}
	\frac{\partial \bm{\tilde{u}}_2(\bm{r},t)}{\partial t}&&=-\frac{1}{(G^2+\alpha^2D^2)^2} \int dt'\int d\bm{r}' \int\frac{d^d\bm{k}}{(2\pi)^d}\int \frac{d\omega}{2\pi} \mathcal{G}(\bm{k},\omega)(-i\omega) e^{i\bm{k}\cdot(\bm{r}-\bm{r'})-i\omega(t-t')}\begin{pmatrix}
		\alpha D&G\\
		-G&\alpha D
	\end{pmatrix}\times\nonumber \\&&\int \frac{d^d \bm{q}_1}{(2\pi)^d}	\begin{pmatrix} 
q_{1x}^2&q_{1x}q_{1y}\\q_{1x}q_{1y}&	q_{1y}^2
		\end{pmatrix}U(\bm{q}_1)e^{i\bm{q}_1\cdot (\bm{r'}+\bm{v_d}t')}  \sum_{\bm{g}_1}\rho(\bm{g}_1)e^{i\bm{g}_1\cdot\bm{r'}}\times\nonumber\\
&&\int dt'' \int d\bm{r''} \int \frac{d^d \bm{k'}}{(2\pi)^d}\int \frac{d\omega'}{2\pi}\mathcal{G}(\bm{k'},\omega') e^{i\bm{k'}\cdot(\bm{r'}-\bm{r''})-i\omega'(t'-t'')}\begin{pmatrix}
	\alpha D&G\\
	-G&\alpha D
\end{pmatrix} \sum_{\bm{g}_2}\rho(\bm{g}_2)e^{i\bm{g}_2\cdot\bm{r''}} \times\nonumber\\
&&\int \frac{d^d \bm{q}_2}{(2\pi)^d}\begin{pmatrix}
	i q_{2x}\\ iq_{2y}
\end{pmatrix} U(\bm{q}_2)e^{i\bm{q}_2\cdot(\bm{r''}+\bm{v_d}t'')}	
\end{eqnarray}

We can take integrals with respect to the space and time, and take the average over disorders, several delta functions would appear on the right-hand side:
\begin{eqnarray}
	\overline{U(\bm{q}_1)U(\bm{q}_2)}&&=(2\pi)^d\Lambda(\bm{q}_2)\delta(\bm{q}_1+\bm{q}_2),\\
	\int dt' e^{i\omega t'} e^{i\bm{q}_1\cdot \bm{v_d} t'}e^{-i\omega't'}&&=2\pi\delta(\omega-\omega'+\bm{q}_1\cdot \bm{v_d}),\\
	\int d\bm{r'} e^{-i\bm{k}\cdot\bm{r'}}e^{i\bm{q}_1\cdot\bm{r'}}e^{i\bm{g}_1\cdot\bm{r'}}e^{i\bm{k'}\cdot\bm{r'}}&&=(2\pi)^d\delta (\bm{k'}-\bm{k}+\bm{q}_1+\bm{g}_1),\\
	\int dt'' e^{i\omega' t''}e^{i\bm{q}_2\cdot\bm{v_d} t''}&&=2\pi\delta (\omega'+\bm{q}_2\cdot\bm{v_d}),\\
	\int d\bm{r''} e^{-i\bm{k'}\cdot\bm{r''}}e^{i\bm{g}_2\cdot\bm{r''}}e^{i\bm{q}_2\cdot\bm{r''}}&&=(2\pi)^d\delta(\bm{g}_2+\bm{q}_2-\bm{k'}).	
\end{eqnarray}
Here, the information of impurity potential and functional profile is encoded in the impurity average function: $\Lambda(\bm{q})$.  If the impurity is short-ranged, $\Lambda(\bm{q})$ is more uniform with respect to $\bm{q}$, while if the impurity function is long-ranged, $\Lambda(\bm{q})$ would be peaked around $\bm{q}=0$. In practice, the impurity size could play a similar role to that of the pinning strength.
 Take the volume average, and consider  constraints from   delta functions: $\bm{q}_2=-\bm{q}_1=\bm{q}$,  $\bm{g}_1=-\bm{g}_2=\bm{g}$, $\omega'=-\bm{q}\cdot\bm{v_d}$,  we find

\begin{eqnarray}
	\left<\frac{\partial \bm{\tilde{u}}_2(\bm{r},t)}{\partial t}\right>&&=\frac{1}{(G^2+\alpha^2D^2)^2} \sum_{\bm{g}} \int \frac{d^d\bm{q}}{(2\pi)^d} |\rho(\bm{g})|^2\Lambda (\bm{q})\begin{pmatrix}
		\alpha D&G\\
		-G&\alpha D
	\end{pmatrix}	\begin{pmatrix} 
q_{x}^2&q_{x}q_{y}\\q_{x}q_{y}&	q_{y}^2
		\end{pmatrix}\times\nonumber\\
&& \text{Im}\left[\mathcal{G}(\bm{q}-\bm{g},-\bm{q}\cdot\bm{v_d})\right]\begin{pmatrix}
	\alpha D&G\\
	-G&\alpha D
\end{pmatrix}\begin{pmatrix}
q_x\\ q_y
\end{pmatrix}.
\end{eqnarray}
Therefore, 
\begin{eqnarray}
\left<\frac{\partial \tilde{\bm{u}}(\bm{r},t)}{\partial t} \right>&&=\left[\frac{1}{G^2+\alpha^2D^2}\begin{pmatrix}
	G^2+\alpha\beta D^2&GD(\beta-\alpha)\\GD(\alpha-\beta)&G^2+\alpha\beta D^2
\end{pmatrix}\begin{pmatrix}
	v_{sx}\\v_{sy}
\end{pmatrix}-\bm{v_d}\right]+\nonumber\\
&&\frac{1}{(G^2+\alpha^2D^2)^2} \sum_{\bm{g}} \int \frac{d^d\bm{q}}{(2\pi)^d} |\rho(\bm{g})|^2\Lambda (\bm{q})\begin{pmatrix}
	G&\alpha D\\ \alpha D&-G
\end{pmatrix}	\begin{pmatrix} 
q_{x}^2&q_{x}q_{y}\\q_{x}q_{y}&	q_{y}^2
		\end{pmatrix}\times\nonumber\\
&& \text{Im}\left[\mathcal{G}(\bm{q}-\bm{g},-\bm{q}\cdot\bm{v_d})\right]\begin{pmatrix}
	G&\alpha D\\ 
	\alpha D& -G
\end{pmatrix}\begin{pmatrix}
	q_x\\ q_y
\end{pmatrix}
\end{eqnarray}
Set $\left<\frac{\partial \tilde{\bm{u}}(\bm{r},t)}{\partial t} \right>=0$, the self-consistent equation for the velocity is 
\begin{eqnarray}
	\bm{v_d}&&=\frac{1}{G^2+\alpha^2D^2}\begin{pmatrix}
		G^2+\alpha\beta D^2&GD(\beta-\alpha)\\GD(\alpha-\beta)&G^2+\alpha\beta D^2
	\end{pmatrix}\begin{pmatrix}
		v_{sx}\\v_{sy}
	\end{pmatrix}+\nonumber\\
	&&\frac{1}{(G^2+\alpha^2D^2)^2} \sum_{\bm{g}} \int \frac{d^d\bm{q}}{(2\pi)^d} |\rho(\bm{g})|^2\Lambda (\bm{q})\begin{pmatrix}
		\alpha D&G\\
		-G&\alpha D
	\end{pmatrix}	\begin{pmatrix} 
q_{x}^2&q_{x}q_{y}\\q_{x}q_{y}&	q_{y}^2
		\end{pmatrix}\times\nonumber\\
	&& \text{Im}\left[\mathcal{G}(\bm{q}-\bm{g},-\bm{q}\cdot\bm{v_d})\right]\begin{pmatrix}
		\alpha D&G\\
		-G&\alpha D
	\end{pmatrix}\begin{pmatrix}
		q_x\\ q_y
	\end{pmatrix}.
\end{eqnarray}

As argued in the main text, the largest imaginary part is contributed by $\bm{k}=\bm{q}-\bm{g}$ in long wave limit ($\bm{k}$ is small). 
To further proceed, let us evaluate $\text{Im}[\mathcal{G}(\bm{k},\omega)]$.

\begin{equation}
		\mathcal{G}(\bm{k},\omega)=\frac{1}{-i\omega-\frac{1}{G^2+(D\alpha)^2}\begin{pmatrix}
		\alpha D&G\\
		-G&\alpha D
	\end{pmatrix}  \mathcal{D}(\bm{k})}=-\frac{G^2+\alpha^2D^2}{\lambda(\bm{k})}i\omega-\frac{\mathcal{D}(\bm{k})}{\lambda(\bm{k})}\begin{pmatrix}
\alpha D&-G\\ G&\alpha D
\end{pmatrix},
\end{equation}
where 
\begin{equation}
	\lambda(\bm{k})=[\mathcal{D}(\bm{k})+\omega(i\alpha D+G)][\mathcal{D}(\bm{k})+\omega(i\alpha D-G)]=\mathcal{D}(\bm{k})^2-(G^2+\alpha^2D^2)\omega^2+2i\alpha D\omega \mathcal{D}(\bm{k}).
\end{equation}
The imaginary part of Green's function is given by
\begin{eqnarray}
\text{Im}[\mathcal{G}(\bm{k},\omega)]&&=-\frac{(G^2+\alpha^2D^2)[\mathcal{D}^2(\bm{k})-(G^2+\alpha^2D^2)\omega^2]\omega}{[\mathcal{D}^2(\bm{k})-(G^2+\alpha^2D^2)\omega^2]^2+4\alpha^2D^2\omega^2\mathcal{D}^2(\bm{k})}\nonumber\\
&&+\frac{2\alpha D\omega \mathcal{D}^2(\bm{k})}{[\mathcal{D}^2(\bm{k})-(G^2+\alpha^2D^2)\omega^2]^2+4\alpha^2D^2\omega^2\mathcal{D}^2(\bm{k})}\begin{pmatrix}
\alpha D&-G\\ G&\alpha D
\end{pmatrix}.
\end{eqnarray}
The largest imaginary part is given by the real mode $\omega_{\bm{k}}= \mathcal{D}(\bm{k})/\sqrt{G^2+\alpha^2D^2}$. As a result, the first term in $\text{Im}[\mathcal{G}(\bm{k},\omega)]$ can be negligible. In 2D,  we can expand 
\begin{equation}
	\mathcal{D}(\bm{k})=K_xk_x^2+K_yk_y^2.
\end{equation}
Now we can show that 
\begin{eqnarray}
&&\int \frac{d^{2}\bm{k}}{(2\pi)^2} \frac{2\alpha D\omega \mathcal{D}^2(\bm{k})}{[\mathcal{D}^2(\bm{k})-(G^2+\alpha^2D^2)\omega^2]^2+4\alpha^2D^2\omega^2\mathcal{D}^2(\bm{k})}\nonumber\\&&=(K_xK_y)^{-1/2}\int_0^{+\infty} \frac{dk' 2\pi k'}{(2\pi)^2} \frac{2\alpha D\omega k'^4}{(k'^4-(G^2+\alpha^2D^2)\omega^2)^2+4\alpha^2D^2\omega^2k'^4}\nonumber\\
&&=\frac{(K_xK_y)^{-1/2}\text{sgn}(\omega)}{4 }.
\end{eqnarray}
where the integral $\int_{0}^{+\infty} dt \frac{t^2}{(t^2-\frac{\sqrt{G^2+\alpha^2D^2}}{2\alpha D})^2+t^2}=\frac{\pi}{2}$ is used with $t=\frac{k'^2}{2\alpha D|\omega|}$. Note that $\alpha D\neq 0$ is taken. 
The multiplications  between matrices give 
\begin{eqnarray}
\begin{pmatrix}
	\alpha D&G\\
	-G&\alpha D
\end{pmatrix}	\begin{pmatrix} 
q_{x}^2&q_{x}q_{y}\\q_{x}q_{y}&	q_{y}^2
		\end{pmatrix}\begin{pmatrix}
\alpha D&-G\\
G&\alpha D
\end{pmatrix}\begin{pmatrix}
\alpha D&G\\
-G&\alpha D
\end{pmatrix}\begin{pmatrix}
q_x\\q_y
\end{pmatrix}&&=(G^2+\alpha^2D^2)(q_x^2+q_y^2)\begin{pmatrix}
Gq_y+\alpha D q_x\\ -Gq_x+\alpha Dq_y
\end{pmatrix}.
\end{eqnarray}
Finally, we obtain
\begin{equation}
\delta v_d=\frac{(K_xK_y)^{-1/2}}{4(G^2+\alpha^2D^2)}\sum_{\bm{g}} |\rho(\bm{g})|^2\Lambda (\bm{g})\text{sgn}(-\bm{g}\cdot \bm{v}_{d0})|\bm{g}|^2\begin{pmatrix}
    Gg_y+\alpha D g_x\\ -Gg_x+\alpha Dg_y
\end{pmatrix}
\end{equation}

We have shown that $\delta v_d$ respects out-of-plane rotational symmetry in the main text, which is inherited from the Theiele equation. Without loss of generality, let us set $v_{d0}$ to be along $x$-direction. In this case,  after summing over the six smallest $\bm{g}$ vectors: $\bm{g}_j=\sqrt{3}\kappa_0(\sin \frac{(j-1)\pi}{3},\cos\frac{(j-1)\pi}{3})$ with $\kappa_0=\frac{4\pi}{3a}$, $j$ are integers from 1 to 6,  we find
\begin{equation}
    \delta \bm{v}_d=\frac{9\kappa_0^3|\rho_1|^2\Lambda_0\alpha D}{4\sqrt{K_xK_y}(G^2+\alpha^2D^2)}\begin{pmatrix}
        -1\\
        \frac{G}{\alpha D}
    \end{pmatrix}
\end{equation}
where $\rho_1= \rho(g_j), \Lambda_0= \Lambda(g_j)$. 
In the 3D case, 
\begin{equation}
\mathcal{D}(\bm{k})=K_xk_x^2+K_yk_y^2+K_zk_z^2.
\end{equation}
Then,
\begin{eqnarray}
&&\int \frac{d^{3}\bm{k}}{(2\pi)^3} \frac{2\alpha D\omega \mathcal{D}^2(\bm{k})}{[\mathcal{D}^2(\bm{k})-(G^2+\alpha^2D^2)\omega^2]^2+4\alpha^2D^2\omega^2\mathcal{D}^2(\bm{k})}\nonumber\\
&&=(K_xK_yK_z)^{-1/2}\int_{0}^{+\infty} \frac{4\pi k'^2 dk'}{(2\pi)^3} \frac{2\alpha D\omega k'^4}{(k'^4-(G^2+\alpha^2D^2)\omega^2)^2+4\alpha^2D^2\omega^2k'^4}\nonumber\\
&&=\frac{(K_xK_yK_z)^{-1/2}\Gamma(\frac{G^2+\alpha^2D^2}{4\alpha^2D^2})\text{sgn}(\omega)\sqrt{2\alpha D|\omega|}}{2\pi^2}.
\end{eqnarray}
where $\Gamma(a)= \int_0^{+\infty} dx \frac{x^6}{(x^4-a)^2+x^4}$.
Similarly, we can obtain
\begin{equation}
\delta v_{d}=\frac{(2\alpha D)^{1/2}(K_xK_yK_z)^{-1/2}\Gamma(\frac{G^2+\alpha^2D^2}{4\alpha^2D^2})}{2\pi^2 (G^2+\alpha^2D^2)} \sum_{\bm{g}} |\rho(\bm{g})|^2\Lambda (\bm{g})\text{sgn}(-\bm{g}\cdot \bm{v}_{d0})\sqrt{|\bm{g}\cdot\bm{v}_{d0}|}|\bm{g}|^2\begin{pmatrix}
    Gg_y+\alpha D g_x\\ -Gg_x+\alpha Dg_y
\end{pmatrix}.
\end{equation}
After summing over the six smallest $\bm{g}$ vectors, we find
\begin{equation}
\delta \bm{v}_d=\frac{9\sqrt{3}\kappa_0^{7/2}\Gamma(\frac{G^2+\alpha^2D^2}{4\alpha^2D^2})|\rho_1|^2\Lambda_0\alpha D\sqrt{\alpha Dv_{d0}}}{\pi^2\sqrt{K_xK_yK_z}(G^2+\alpha^2D^2)}\begin{pmatrix}
-1\\ \frac{G}{\alpha D}
\end{pmatrix}.
\end{equation}

\section{Helical spin order case}

In this section, we consider the helical spin order case. Without loss of generality, we denote the helical spin order has a variation along $x$-direction. The Thiele equation for the helical spin order would be
\begin{equation}
	D(\beta v_{s}-\alpha v_{d})+F=0.
\end{equation}
For simplicity, we have omitted the index $x$ in the following. 
The equation of motion becomes
\begin{equation}
	\frac{\partial u(\bm{r},t)}{\partial t}=\frac{\beta}{\alpha}v_s+\frac{1}{\alpha D} \int d\bm{r'} \mathcal{D}(\bm{r}-\bm{r'})u(\bm{r'},t')+\frac{1}{\alpha D} f_{imp}(\bm{r}+u(\bm{r},t))\rho(\bm{r}).
\end{equation}
Similarly, by defining $u(\bm{r},r)=v_{d}t+\tilde{u}(\bm{r},t)$, the equation of motion can be rewritten as
\begin{equation}
	[\frac{\partial}{\partial t}-\frac{1}{\alpha D}\int d\bm{r'} \mathcal{D}(\bm{r}-\bm{r'})]\tilde{u}(\bm{r'},t)=\frac{\beta}{\alpha}v_s+\frac{1}{\alpha D}f_{imp}(\bm{r}+u(\bm{r},t))\rho(\bm{r}).
\end{equation}
Let us define the Green's function $\mathcal{G}(\bm{r},t)$ so that
\begin{equation}
		[\frac{\partial}{\partial t}-\frac{1}{\alpha D}\int d\bm{r'} \mathcal{D}(\bm{r}-\bm{r'})]\tilde{u}(\bm{r'},t)\mathcal{G}(\bm{r'},t)=\delta(\bm{r})\delta(t).
\end{equation}
It is easy to obtain
\begin{equation}
	\mathcal{G}^{-1}(\bm{k},\omega)=-i\omega-\frac{\mathcal{D}(\bm{k})}{\alpha D}.
\end{equation}
The displacement $\tilde{u}(\bm{r},t)$ is given by
\begin{equation}
	\tilde{u}(\bm{r},t)=\int d\bm{r'}\int dt' \mathcal{G}(\bm{r}-\bm{r'},t-t')[\frac{\beta}{\alpha}v_s-v_d+\frac{1}{\alpha D} f_{imp}(\bm{r'}+u(\bm{r'},t'))\rho(\bm{r'})].
\end{equation} 
Then up to the second order,
\begin{eqnarray}
	\tilde{u}_0(\bm{r},t)&&=\int d\bm{r'} dt' \mathcal{G}(\bm{r}-\bm{r'},t-t')[\frac{\beta}{\alpha}v_s-v_d],\\
	\tilde{u}_1(\bm{r},t)&&=\frac{1}{\alpha D} \int d\bm{r'} \int dt' \mathcal{G}(\bm{r}-\bm{r'},t-t') f_{imp}(\bm{r'}+v_d t')\rho(\bm{r'}),\\
	\tilde{u}_2(\bm{r},t)&&=\frac{1}{\alpha D} \int d\bm{r'}\int dt' \mathcal{G}(\bm{r}-\bm{r'},t-t')\partial_x f_{imp}(\bm{r'}+v_dt')\tilde{u}_1(\bm{r'},t')\rho(\bm{r'}).
\end{eqnarray}

Following a similar procedure in Sec. II, using $\braket{\frac{\partial u(\bm{r},t)}{\partial t}}=0$, we obtain
\begin{equation}
	v_d=v_{d0}+\frac{1}{\alpha^2D^2 }\sum_{\bm{g}}|\rho(\bm{g})|^2\int\frac{d^d\bm{q}}{(2\pi)^d}\Lambda(\bm{q})q_x^3\text{Im}[\mathcal{G}(\bm{q}-\bm{g},-q_xv_d)].
\end{equation}
where the intrinsic drift velocity $v_{d0}=\frac{\beta}{\alpha}v_s$. Consider the $\text{Im}[G(\bm{k},\omega)]$ is dominant in the long wave limit ($\bm{k}\rightarrow 0$) and expand $\mathcal{D}(\bm{k})=K_x k_x^2+K_yk_y^2$  in $d=2$, we find
\begin{equation}
	v_d\approx v_{d0}-\frac{(K_xK_y)^{-1/2}}{8\alpha D}\sum_{\bm{g}}|\rho(\bm{g})|^2\Lambda(\bm{g})|g_x|^3 
\end{equation}
and similarly in $d=3$ case, $\mathcal{D}(\bm{k})=K_x k_x^2+K_yk_y^2+K_zk_z^2$,  we obtain
\begin{equation}
	v_d\approx v_{d0}-\frac{(K_xK_yK_z)^{-1/2}}{4\sqrt{2}\pi(\alpha D)^{1/2}}\sum_{\bm{g}}|\rho(\bm{g})|^2\Lambda(\bm{g})|g_x|^3\sqrt{|g_x v_{d0}|}.
\end{equation}

After summing over the smallest  reciprocal lattice vectors for the helix: $\bm{g}=(\pm 1,0)g_0$ with $g_0=\frac{\pi}{a}$, the correction 
\begin{equation}
	\Delta v_{d}=\frac{\beta}{\alpha} v_s-\chi^{HL}_d (v_{d0})^{\frac{d-2}{2}}
\end{equation}
where
\begin{equation}
\chi_d^{HL}=\begin{cases}
\frac{(K_xK_y)^{-1/2}|\rho_1|^2\Lambda_0g_0^3}{4\alpha D}, & \text{for } d=2 \\
\frac{(K_xK_yK_z)^{-1/2}|\rho_1|^2\Lambda_0g_0^{7/2}}{2\sqrt{2}\pi(\alpha D)^{1/2}},& \text{for } d=3
\end{cases}	
\end{equation}
Here, we have set $\rho(g_0)=\rho_1$ and $\Lambda(g_0)=\Lambda_0$ in this case.

\section{Numerical  method details}
\subsection{Details for the main text Fig.2}

The main text Fig.2 is obtained from the main text Eq.~(6) and (9) by numerically integrating $\bm{q}$  and summing over the smallest reciprocal lattice $\bm{g}$ vectors.  For the SkX, the $4000\times 4000$ in-plane momentum grids of $\bm{q}$ are taken with a hexagonal boundary (the boundary length is $4\kappa_0$); while for the HL, the $1000\times 1000$ in-plane momentum grids are taken with a square boundary (the boundary length is $4g_0$). In the 3D case, the 1000 out-of-plane momentum points of $\bm{q}$ within $[-2 g_0, 2g_0]$ are used for both SkX and HL in evaluating the integral.    Also, we set the elastic coefficients $K_j=10000$, the lattice constant $a$ as a natural unit of one, the damping parameter $\alpha=0.2$, the dissipative coefficient $D=5\pi$, the additional parameter $G=\pm 4\pi$ for the SkX.
\subsection{Micromagnetic simulations}
The micromagnetic simulations were performed using MuMax3~\cite{vansteenkiste_design_2014}.
The Landau-Lifshitz-Gilbert (LLG) equation is numerically solved
\begin{equation}
  \dot{\bm{s}} = -|\gamma_{0}|\bm{s} \times \bm{H}_{\text{eff}} + \alpha \bm{s} \times \dot{\bm{s}} + \frac{p a^3}{2eS} (\bm{j}_s\cdot \bold{\nabla}) \bm{s} - \frac{pa^3 \beta}{2eM_s^2}\bm{s}\times(\bm{j}_s \cdot \bold{\nabla}) \bm{s},
\end{equation}
where $\textbf{s}$ is the unit vector of spin, $\gamma_{0}$ is the gyromagnetic constant, $\alpha$ is the Gilbert damping constant, $M_s$ is the saturation magnetization, and $\bm{H}_{\text{eff}} = -\frac{1}{\mu_{0}M_{s}}\frac{\delta \mathcal{H}}{\delta \bm{S}}$ is the effective field. 
The spin transfer torque effect of the current is described by the last two terms~\cite{slonczewski_current_driven_1996, berger_emission_1996, zhang_roles_2004}.
$\beta$ describes the non-adiabaticity of the spin transfer torque effect.
The current is applied along the x-direction in the simulations.

A typical chiral magnet can be described by the following Hamiltonian density
\begin{equation}
  \mathcal{H} = A(\nabla \bm{S})^2 - D\bm{S}\cdot (\nabla \times \bm{S}) - \mu_{0}M_{s}\bm{B}\cdot\bm{S} 
\end{equation}
The corresponding parameters and their values employed in the simulations are: the saturation magnetization $M_s = 111 $ kA/m, the exchange stiffness $A=3.645$ pJ/m, and the Dzyaloshinskii-Moriya interaction strength $D=0.567 \text{ mJ/m}^2$.
An external magnetic field $\bm{B}=0.3$ T (with its direction perpendicular to the skyrmion plane) is used in the simulations to stabilize the skyrmions.
The simulations for the helical state are performed at zero-field.
The cell size is 1 nm $\times$  1 nm $\times$ 1 nm.

We consider magnetic impurities with uniaxial magnetic anisotropy $\mathcal{H}_{imp}=- K_{imp} S_z^2$, where the easy-axis is perpendicular to the skyrmion 2D plane.
For the weak impurity case, an impurity concentration $x=0.1\%$ and impurity strength $K_{imp}=0.2A/l^2$ ($l$ is the cell size) were used.
The simulation results are averaged over 100 impurity distributions.
The skyrmion velocity is extracted by using the emergent electric field method~\cite{koshibae_theory_2018}.
For each current density, the emergent electric field is also averaged over 100 time steps in order to get the skyrmion velocity.
For the transverse correction, the damping value $\alpha=0.2$ is employed in the main text Figs. 3(a) and (b) for computational efficiency, as using smaller damping values results in significantly longer simulation times to obtain a reasonable correction along the transverse direction.

\section{Pinning effects near the skyrmion-helix transition}
\begin{figure}
	\centering
	\includegraphics[width=0.3\linewidth]{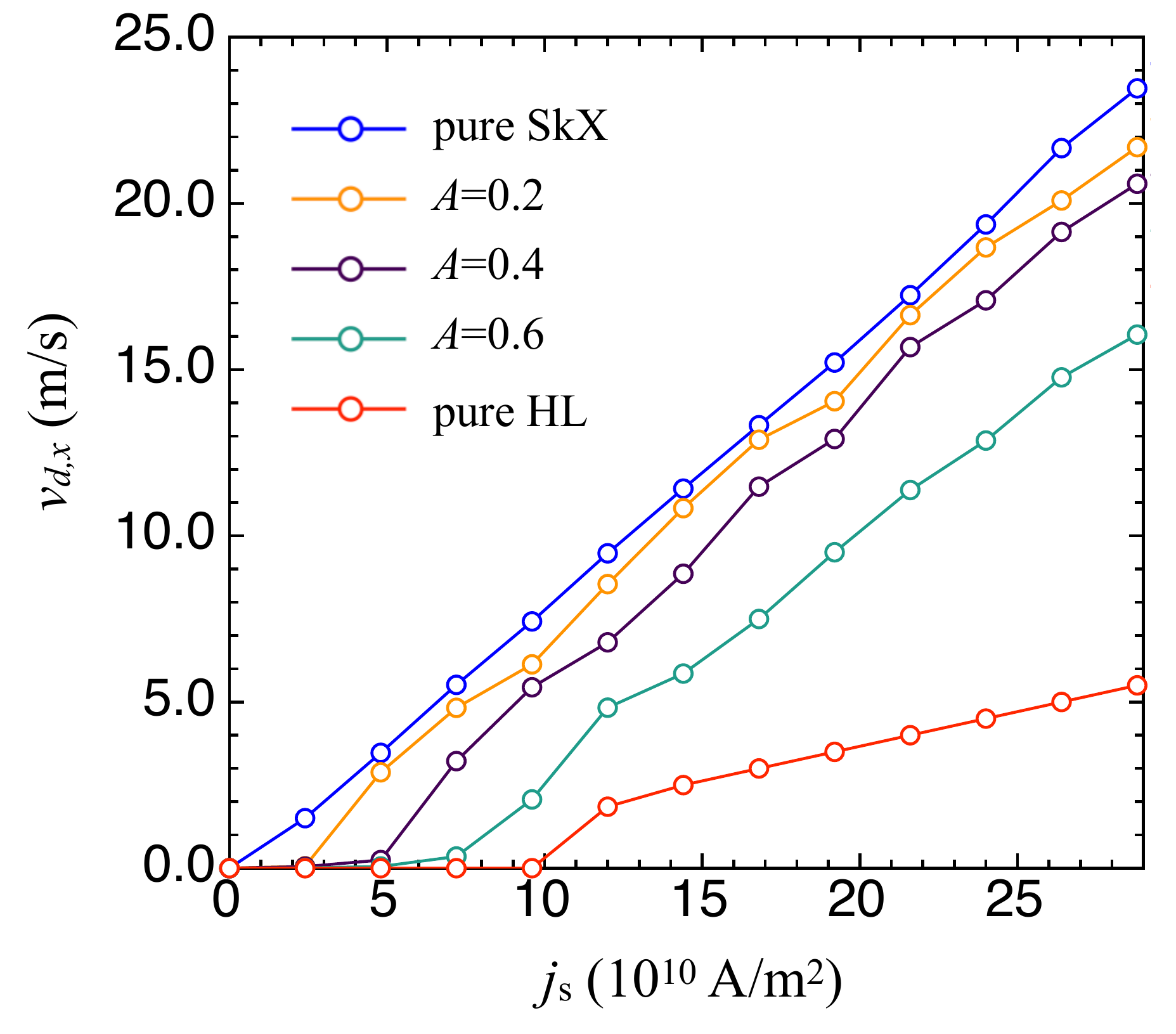}
 	\caption{Simulated longitudinal drift velocity as a function of current density for spin textures with different ratios between the spin helix and skyrmion crystal.}
  \label{fig:figs1}
\end{figure}

In real materials, we may expect some mixed SkX and HL domains near the SkX-HL phase transition boundaries. 
Because the HL exhibits a stronger pinning effect, when the current is larger than the threshold value of depinning SkX, the HL domain area would act like pinning centers to the SkX area. 
As a result, the SkX pinning is expected to be stronger near SkX-HL phase transitions. 
To test this scenario, micromagnetic simulations near the SkX-HL transition boundary were performed. 
A specific magnetic field (0.12 T here) is employed in the simulations so that the energies of SkX and HL are close to each other, and their co-existence phase can be stabilized. 
As shown in Fig.~\ref{fig:figs1}, as the HL domain area ratio ($A=A_{HL}/A_{total}$) increases, the depinning current also gradually increases. 
In addition, the slope of the velocity-current density curve also changes slightly due to the mixing of SkX and HL. 
It should be noted that due to the limited size of the simulated system, the depinning behavior may not be fully captured. 
Many factors can affect the corresponding results such as the orientations of the HL domains, the distribution of the skyrmions, the boundary condition, etc. 
Nevertheless, we expect the overall trend that the critical depinning current increases should be consistent for most of the cases.

\section{Comparison between the skyrmion crystal and a single skyrmion}
\begin{figure}
	\centering
	\includegraphics[width=0.6\linewidth]{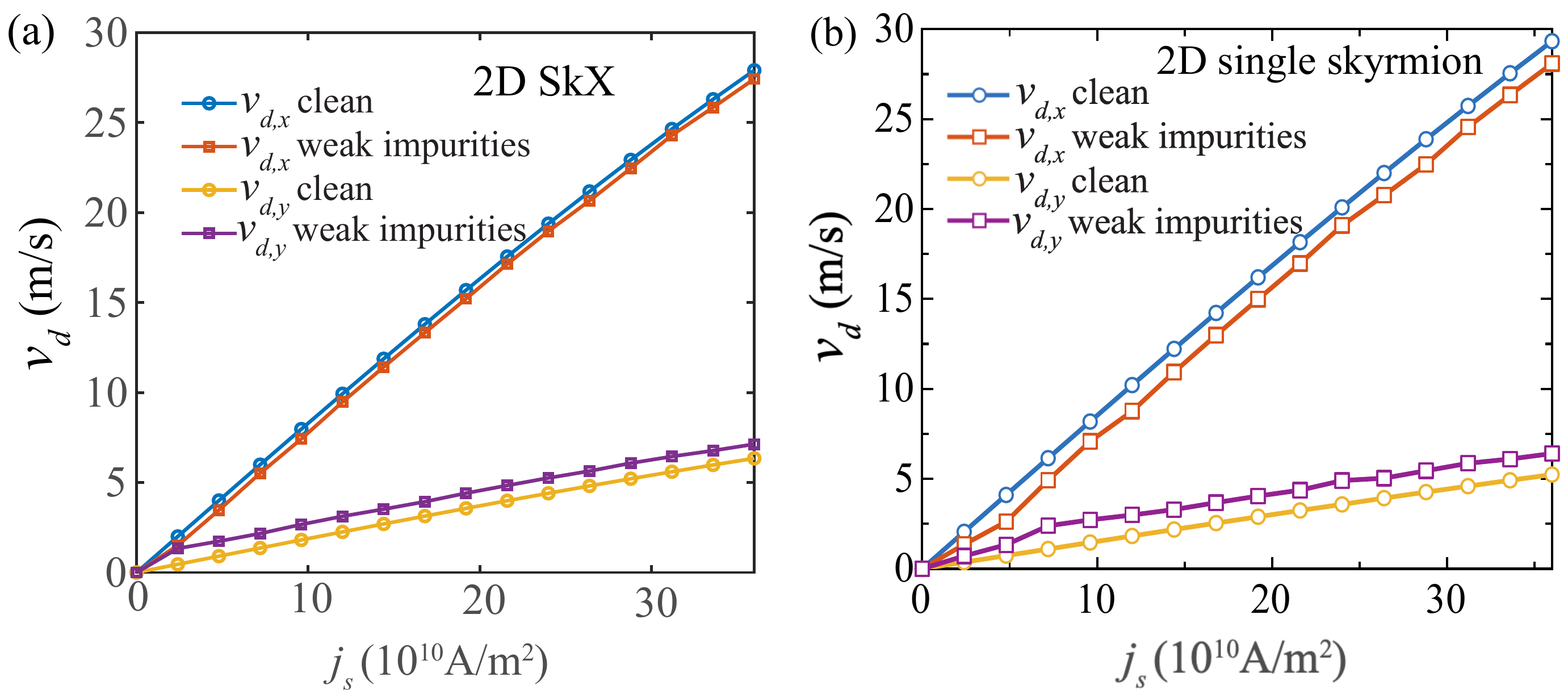}
 	\caption{Drift velocities of a SkX (a) and a single skyrmion (b) as functions of the driving current density.}
  \label{fig:figs2}
\end{figure}

In this section, we discuss the difference between the SkX and a single skyrmion in terms of the velocity correction in the presence of impurities.
To compare these two limits, we show the SkX case and single skyrmion case in Fig.~\ref{fig:figs2} (a) and (b), respectively. 
These two limits are similar in the sense that both cases respect the universal linear current-drift velocity relation, being expected from the Thiele equation. 
The main difference between the two cases is that the ratio of the transverse velocity correction to the longitudinal velocity correction, $\Delta v{d,\perp}/\Delta v_{d,\parallel}$. 
As discussed in the main text, in the case of a SkX, following Eq. 7, this ratio is about $G/\alpha D$, which is a result of both the topology of the skyrmion and the elasticity of the SkX.
In contrast, the transverse velocity correction and the longitudinal velocity correction of a single skyrmion do not follow this trend, which is due to the free motion in the absence of neighboring skyrmions. 

\section{Drift velocity of skyrmion crystal under different impurity strength}
\begin{figure}
	\centering
	\includegraphics[width=0.35\linewidth]{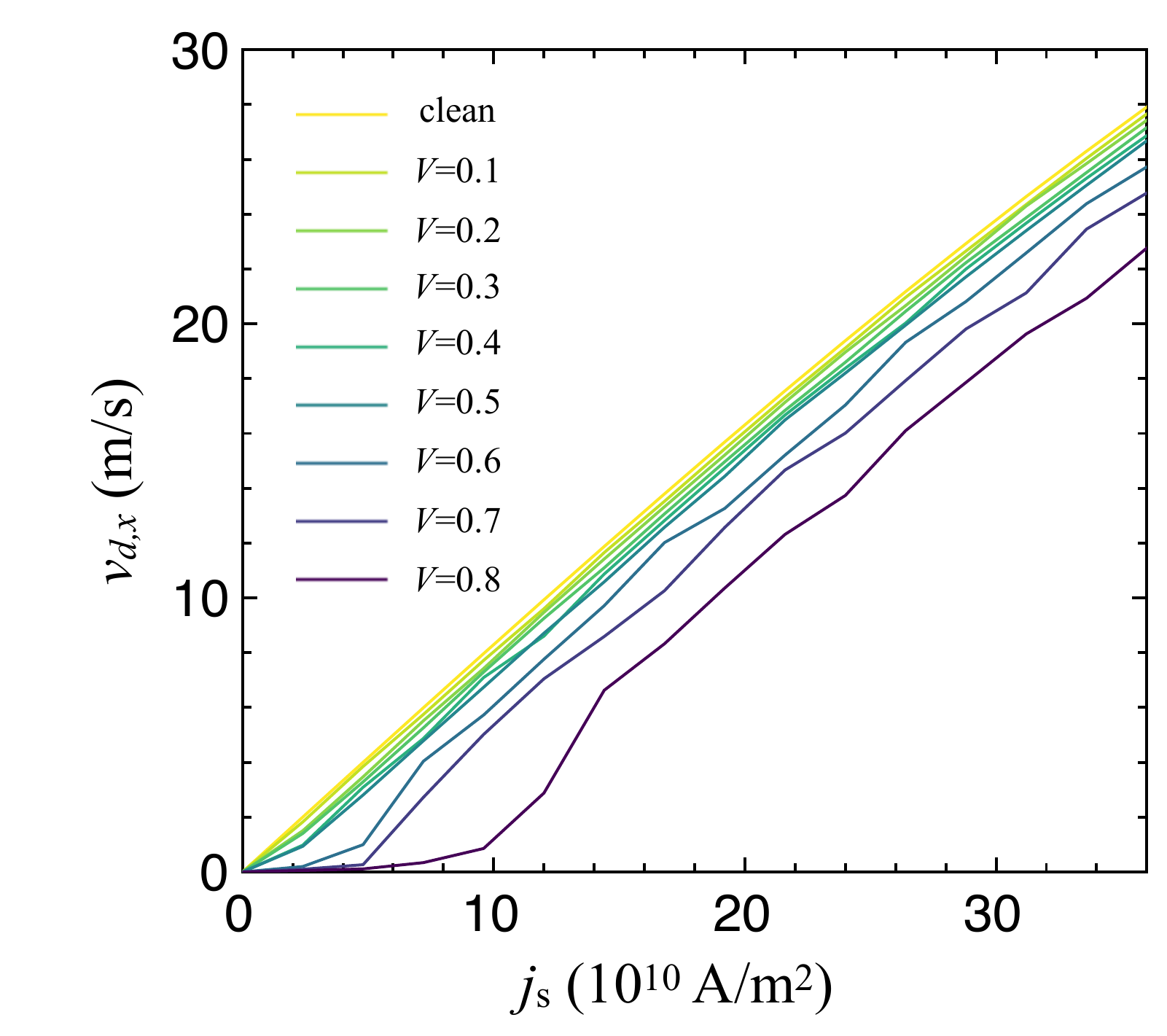}
 	\caption{The simulated velocity-current density curve for a SkX under various impurity strength $V$.}
  \label{fig:figs4}
\end{figure}

In this section, we show the results of the SkX's drift velocity under various impurity strengths $V=|K_{imp} l^2 /A|$.
As shown in Fig.~\ref{fig:figs4}, for $V\le~0.5$, the depinning current does not vary too much within our simulation resolution. 
For $V>~0.5$, the depinning current starts to increase, and the velocity-current curve strongly deviates from that in the clean limit, which suggests a strong disorder regime. 
Our theory assumes that the SkX crystal is only slightly distorted and each skyrmion is a rigid body, so it does not work in the strong disorder regime.

\section{Validation of the Thiele equation}
\begin{figure}
	\centering
	\includegraphics[width=0.6\linewidth]{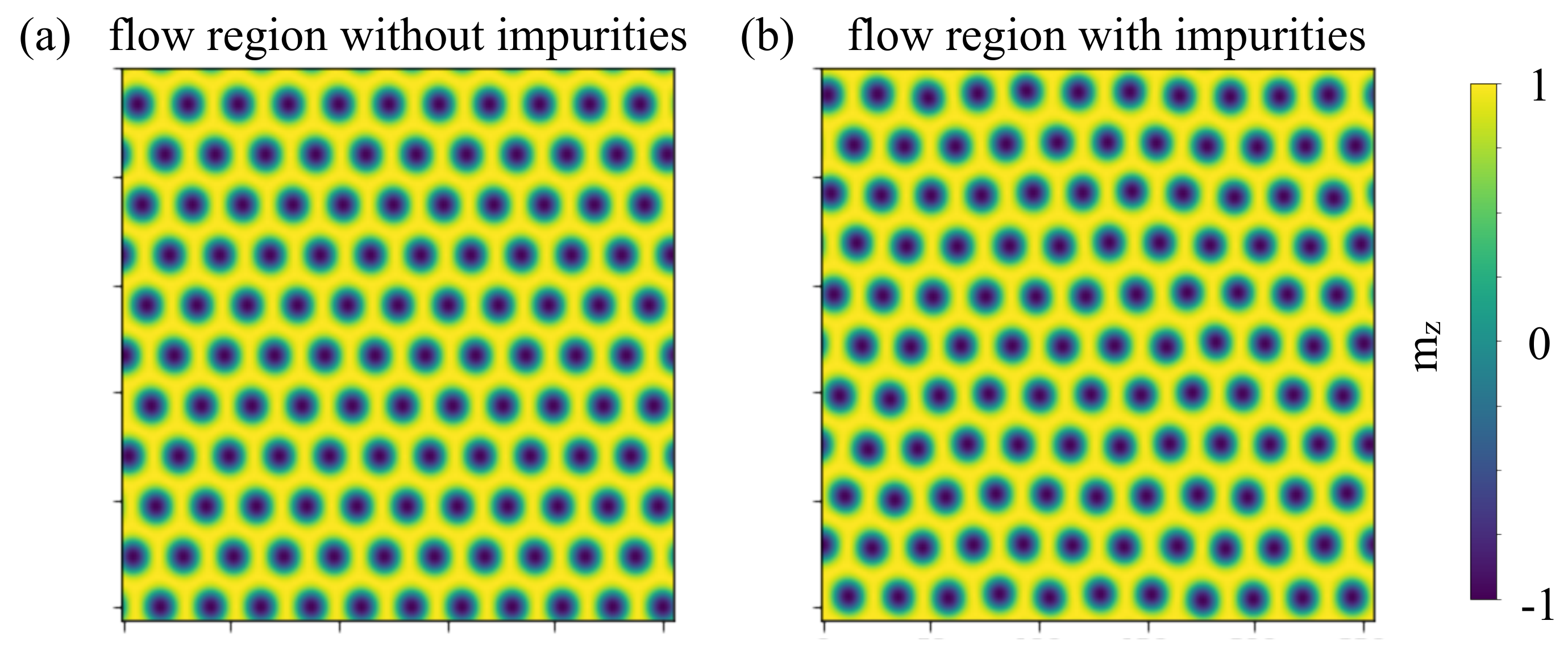}
 	\caption{Real-space snapshots of a flowing skyrmion crystal in a system without impurities (a) and with impurities (b). The strength of the impurity is the same with that used in the main text. The size of the system is 768 nm by 768 nm. The color encodes the z-component of the magnetization.}
  \label{fig:figs5}
\end{figure}

 The Thiele equation approach can usually be applied if the deformation within each skyrmion can be negligible. 
 Indeed, in our consideration, the SkX in chiral magnets behaves as crystal flowing and the deformation of each skyrmion is tiny in the flowing limit. %
 In this case, the center of mass movement is more relevant. 
 Fig. ~\ref{fig:figs5} shows the real-space snapshots of a flowing SkX in a system without impurities (a) and with impurities (b).
It is obvious that the deformation is negligible for each individual skyrmion and only a slight distortion of the SkX can be identified.
 Hence, the approximation for the Thiele equation is justified in our scenario.

 \end{document}